%% file: paper.tex
\documentclass{vldb}
\usepackage{amsmath}
\usepackage{hyperref}
\usepackage{algorithm}
\usepackage[noend]{algpseudocode}
\usepackage{balance}  % for  \balance command ON LAST PAGE  (only there!)
\usepackage{subfigure}
\usepackage{xcolor}
\usepackage{multirow}
\usepackage{graphicx}

\newcommand{\squishlist}{
 \begin{list}{$\bullet$}
  { \setlength{\itemsep}{0pt}
     \setlength{\parsep}{0pt}
     \setlength{\topsep}{0pt}
     \setlength{\partopsep}{0pt}
     \setlength{\leftmargin}{1.2em}
     \setlength{\labelwidth}{0.5em}
     \setlength{\labelsep}{0.5em} } }
\newcommand{\squishend}{\end{list}}

\newtheorem{lemma}{Lemma}%[section]
\newtheorem{theorem}{Theorem}%[section]
\newtheorem{proposition}{Proposition}%[section]
\newtheorem{corollary}{Corollary}%[section]
\newtheorem{definition}{Definition}%[section]
\newtheorem{example}{Example}%[section]

\begin{document}

\title{Learning Tuple Probabilities}

\numberofauthors{2}
\author{
\alignauthor
Maximilian Dylla\\
       \affaddr{Max-Planck-Institute for Informatics}\\
       \affaddr{Saarbr\"ucken, Germany}\\
       \email{mdylla@mpi-inf.mpg.de}
\alignauthor
Martin Theobald\\
       \affaddr{University of Ulm}\\
       \affaddr{Ulm, Germany}\\
       \email{martin.theobald@uni-ulm.de}
}

\maketitle

\begin{abstract}
Learning the parameters of complex probabilistic-relational models from labeled
training data is a standard technique in machine learning, which has been
intensively studied in the subfield of Statistical Relational Learning
(SRL), but---so far---this is still an under-investigated topic in the context
of Probabilistic Databases (PDBs). In this paper, we focus on learning the
probability values of base tuples in a PDB from labeled lineage formulas. 
The resulting learning problem can be viewed as the inverse problem to confidence computations in PDBs: given a set of labeled query answers, learn the probability values of the base tuples, such that the marginal probabilities of the query answers again yield in the assigned probability labels.
We analyze the learning problem from a theoretical perspective, cast it into an optimization problem, and provide an algorithm based on stochastic gradient descent. 
Finally, we conclude by an experimental
evaluation on three real-world and one synthetic dataset, thus comparing our approach to various techniques from SRL, reasoning in information extraction, and optimization.
\end{abstract}

\input{intro}
\input{relatedWork}
\input{model}
\input{problem}
\input{solving}
\input{algorithms}
\input{relatedProblems}
\input{experiments}
\input{conclusions}

\small
\bibliographystyle{abbrv}
\bibliography{references}

% ensure same length columns on last page (might need two sub-sequent latex runs)
\balance
\end{document}

\title{Learning Tuple Confidences in Probabilistic Databases - Supplementary Material}
\author{Maximilian Dylla \and Martin Theobald}
\maketitle

This document contains additional material regarding the experiments of the
``Learning Tuple Confidences in Probabilistic Databases" paper. The sections
follow the flow of the paper, where experiments not included in the paper are
last. By relation names we refer to their respective sets of tuples.

\section{Quality Task: SRL Data}
\label{sec:uwcse}
We set $\mathcal{T} := \mathit{CourseLevel}
\cup \mathit{HasPosition} \cup \mathit{InPhase} \cup \mathit{Professor} \cup
\mathit{ProjectMember} \cup \mathit{Publication} \cup \mathit{Student} \cup
\mathit{TaughtBy} \cup \mathit{Ta} \cup \mathit{YearsInProgram} \cup \mathit{Rules}$ and
$\mathcal{T}_l := \mathit{Rules}$. Here, {\em Rules} is the only uncertain relation.
The 49 automatically created rules are:
\[
\begin{array}{l}
\mathit{AdvisedBy}(P_1,P_2) \leftarrow \left( \begin{array}{l} \mathit{Ta}(D,C,P_1,\mathit{Te}) \land \\ \mathit{Ta}(D,C,P_2,\mathit{Te}) \land \\ \end{array} \mathit{Rules}(0) \right) \\
\mathit{AdvisedBy}(P_1,P_2) \leftarrow \left( \begin{array}{l} \mathit{Ta}(D,C,P_1,\mathit{Te}) \land \\ \mathit{TaughtBy}(D,C,P_2,\mathit{Te}) \land \\ \end{array} \mathit{Rules}(1) \right) \\
\mathit{AdvisedBy}(P_1,P_2) \leftarrow \left( \begin{array}{l} \mathit{Ta}(D,C,P_1,\mathit{Te}) \land \\ \mathit{Publication}(D,\mathit{Ti},P_2) \land \\ \end{array} \mathit{Rules}(2) \right) \\
\mathit{AdvisedBy}(P_1,P_2) \leftarrow \left( \begin{array}{l} \mathit{Ta}(D,C,P_1,\mathit{Te}) \land \\ \mathit{YearsInProgram}(D,P_2,Y) \land \\ \end{array} \mathit{Rules}(3) \right) \\
\mathit{AdvisedBy}(P_1,P_2) \leftarrow \left( \begin{array}{l} \mathit{Ta}(D,C,P_1,\mathit{Te}) \land \\ \mathit{HasPosition}(D,P_2,\mathit{Po}) \land \\ \end{array} \mathit{Rules}(4) \right) \\
\mathit{AdvisedBy}(P_1,P_2) \leftarrow \left( \begin{array}{l} \mathit{Ta}(D,C,P_1,\mathit{Te}) \land \\ \mathit{InPhase}(D,P_2,\mathit{Ph}) \land \\ \end{array} \mathit{Rules}(5) \right) \\
\mathit{AdvisedBy}(P_1,P_2) \leftarrow \left( \begin{array}{l} \mathit{Ta}(D,C,P_1,\mathit{Te}) \land \\ \mathit{ProjectMember}(D,\mathit{Pr},P_2) \land \\ \end{array} \mathit{Rules}(6) \right) \\
\end{array}
\]
\[
\begin{array}{l}
\mathit{AdvisedBy}(P_1,P_2) \leftarrow \left( \begin{array}{l} \mathit{TaughtBy}(D,C,P_1,\mathit{Te}) \land \\ \mathit{Ta}(D,C,P_2,\mathit{Te}) \land \\ \end{array} \mathit{Rules}(7) \right) \\
\mathit{AdvisedBy}(P_1,P_2) \leftarrow \left( \begin{array}{l} \mathit{TaughtBy}(D,C,P_1,\mathit{Te}) \land \\ \mathit{TaughtBy}(D,C,P_2,\mathit{Te}) \land \\ \end{array} \mathit{Rules}(8) \right) \\
\mathit{AdvisedBy}(P_1,P_2) \leftarrow \left( \begin{array}{l} \mathit{TaughtBy}(D,C,P_1,\mathit{Te}) \land \\ \mathit{Publication}(D,\mathit{Ti},P_2) \land \\ \end{array} \mathit{Rules}(9) \right) \\
\mathit{AdvisedBy}(P_1,P_2) \leftarrow \left( \begin{array}{l} \mathit{TaughtBy}(D,C,P_1,\mathit{Te}) \land \\ \mathit{YearsInProgram}(D,P_2,Y) \land \\ \end{array} \mathit{Rules}(10) \right) \\
\mathit{AdvisedBy}(P_1,P_2) \leftarrow \left( \begin{array}{l} \mathit{TaughtBy}(D,C,P_1,\mathit{Te}) \land \\ \mathit{HasPosition}(D,P_2,\mathit{Po}) \land \\ \end{array} \mathit{Rules}(11) \right) \\
\mathit{AdvisedBy}(P_1,P_2) \leftarrow \left( \begin{array}{l} \mathit{TaughtBy}(D,C,P_1,\mathit{Te}) \land \\ \mathit{InPhase}(D,P_2,\mathit{Ph}) \land \\ \end{array} \mathit{Rules}(12) \right) \\
\mathit{AdvisedBy}(P_1,P_2) \leftarrow \left( \begin{array}{l} \mathit{TaughtBy}(D,C,P_1,\mathit{Te}) \land \\ \mathit{ProjectMember}(D,\mathit{Pr},P_2) \land \\ \end{array} \mathit{Rules}(13) \right) \\
\mathit{AdvisedBy}(P_1,P_2) \leftarrow \left( \begin{array}{l} \mathit{Publication}(D,\mathit{Ti},P_1) \land \\ \mathit{Ta}(D,C,P_2,\mathit{Te}) \land \\ \end{array} \mathit{Rules}(14) \right) \\
\mathit{AdvisedBy}(P_1,P_2) \leftarrow \left( \begin{array}{l} \mathit{Publication}(D,\mathit{Ti},P_1) \land \\ \mathit{TaughtBy}(D,C,P_2,\mathit{Te}) \land \\ \end{array} \mathit{Rules}(15) \right) \\
\mathit{AdvisedBy}(P_1,P_2) \leftarrow \left( \begin{array}{l} \mathit{Publication}(D,\mathit{Ti},P_1) \land \\ \mathit{Publication}(D,\mathit{Ti},P_2) \land \\ \end{array} \mathit{Rules}(16) \right) \\
\mathit{AdvisedBy}(P_1,P_2) \leftarrow \left( \begin{array}{l} \mathit{Publication}(D,\mathit{Ti},P_1) \land \\ \mathit{YearsInProgram}(D,P_2,Y) \land \\ \end{array} \mathit{Rules}(17) \right) \\
\mathit{AdvisedBy}(P_1,P_2) \leftarrow \left( \begin{array}{l} \mathit{Publication}(D,\mathit{Ti},P_1) \land \\ \mathit{HasPosition}(D,P_2,\mathit{Po}) \land \\ \end{array} \mathit{Rules}(18) \right) \\
\mathit{AdvisedBy}(P_1,P_2) \leftarrow \left( \begin{array}{l} \mathit{Publication}(D,\mathit{Ti},P_1) \land \\ \mathit{InPhase}(D,P_2,\mathit{Ph}) \land \\ \end{array} \mathit{Rules}(19) \right) \\
\mathit{AdvisedBy}(P_1,P_2) \leftarrow \left( \begin{array}{l} \mathit{Publication}(D,\mathit{Ti},P_1) \land \\ \mathit{ProjectMember}(D,\mathit{Pr},P_2) \land \\ \end{array} \mathit{Rules}(20) \right) \\
\mathit{AdvisedBy}(P_1,P_2) \leftarrow \left( \begin{array}{l} \mathit{YearsInProgram}(D,P_1,Y) \land \\ \mathit{Ta}(D,C,P_2,\mathit{Te}) \land \\ \end{array} \mathit{Rules}(21) \right) \\
\mathit{AdvisedBy}(P_1,P_2) \leftarrow \left( \begin{array}{l} \mathit{YearsInProgram}(D,P_1,Y) \land \\ \mathit{TaughtBy}(D,C,P_2,\mathit{Te}) \land \\ \end{array} \mathit{Rules}(22) \right) \\
\mathit{AdvisedBy}(P_1,P_2) \leftarrow \left( \begin{array}{l} \mathit{YearsInProgram}(D,P_1,Y) \land \\ \mathit{Publication}(D,\mathit{Ti},P_2) \land \\ \end{array} \mathit{Rules}(23) \right) \\
\mathit{AdvisedBy}(P_1,P_2) \leftarrow \left( \begin{array}{l} \mathit{YearsInProgram}(D,P_1,Y) \land \\ \mathit{YearsInProgram}(D,P_2,Y) \land \\ \end{array} \mathit{Rules}(24) \right) \\
\mathit{AdvisedBy}(P_1,P_2) \leftarrow \left( \begin{array}{l} \mathit{YearsInProgram}(D,P_1,Y) \land \\ \mathit{HasPosition}(D,P_2,\mathit{Po}) \land \\ \end{array} \mathit{Rules}(25) \right) \\
\mathit{AdvisedBy}(P_1,P_2) \leftarrow \left( \begin{array}{l} \mathit{YearsInProgram}(D,P_1,Y) \land \\ \mathit{InPhase}(D,P_2,\mathit{Ph}) \land \\ \end{array} \mathit{Rules}(26) \right) \\
\end{array}
\]
\[
\begin{array}{l}
\mathit{AdvisedBy}(P_1,P_2) \leftarrow \left( \begin{array}{l} \mathit{YearsInProgram}(D,P_1,Y) \land \\ \mathit{ProjectMember}(D,\mathit{Pr},P_2) \land \\ \end{array} \mathit{Rules}(27) \right) \\
\mathit{AdvisedBy}(P_1,P_2) \leftarrow \left( \begin{array}{l} \mathit{HasPosition}(D,P_1,\mathit{Po}) \land \\ \mathit{Ta}(D,C,P_2,\mathit{Te}) \land \\ \end{array} \mathit{Rules}(28) \right) \\
\mathit{AdvisedBy}(P_1,P_2) \leftarrow \left( \begin{array}{l} \mathit{HasPosition}(D,P_1,\mathit{Po}) \land \\ \mathit{TaughtBy}(D,C,P_2,\mathit{Te}) \land \\ \end{array} \mathit{Rules}(29) \right) \\
\mathit{AdvisedBy}(P_1,P_2) \leftarrow \left( \begin{array}{l} \mathit{HasPosition}(D,P_1,\mathit{Po}) \land \\ \mathit{Publication}(D,\mathit{Ti},P_2) \land \\ \end{array} \mathit{Rules}(30) \right) \\
\mathit{AdvisedBy}(P_1,P_2) \leftarrow \left( \begin{array}{l} \mathit{HasPosition}(D,P_1,\mathit{Po}) \land \\ \mathit{YearsInProgram}(D,P_2,Y) \land \\ \end{array} \mathit{Rules}(31) \right) \\
\mathit{AdvisedBy}(P_1,P_2) \leftarrow \left( \begin{array}{l} \mathit{HasPosition}(D,P_1,\mathit{Po}) \land \\ \mathit{HasPosition}(D,P_2,\mathit{Po}) \land \\ \end{array} \mathit{Rules}(32) \right) \\
\mathit{AdvisedBy}(P_1,P_2) \leftarrow \left( \begin{array}{l} \mathit{HasPosition}(D,P_1,\mathit{Po}) \land \\ \mathit{InPhase}(D,P_2,\mathit{Ph}) \land \\ \end{array} \mathit{Rules}(33) \right) \\
\mathit{AdvisedBy}(P_1,P_2) \leftarrow \left( \begin{array}{l} \mathit{HasPosition}(D,P_1,\mathit{Po}) \land \\ \mathit{ProjectMember}(D,\mathit{Pr},P_2) \land \\ \end{array} \mathit{Rules}(34) \right) \\
\mathit{AdvisedBy}(P_1,P_2) \leftarrow \left( \begin{array}{l} \mathit{InPhase}(D,P_1,\mathit{Ph}) \land \\ \mathit{Ta}(D,C,P_2,\mathit{Te}) \land \\ \end{array} \mathit{Rules}(35) \right) \\
\mathit{AdvisedBy}(P_1,P_2) \leftarrow \left( \begin{array}{l} \mathit{InPhase}(D,P_1,\mathit{Ph}) \land \\ \mathit{TaughtBy}(D,C,P_2,\mathit{Te}) \land \\ \end{array} \mathit{Rules}(36) \right) \\
\mathit{AdvisedBy}(P_1,P_2) \leftarrow \left( \begin{array}{l} \mathit{InPhase}(D,P_1,\mathit{Ph}) \land \\ \mathit{Publication}(D,\mathit{Ti},P_2) \land \\ \end{array} \mathit{Rules}(37) \right) \\
\mathit{AdvisedBy}(P_1,P_2) \leftarrow \left( \begin{array}{l} \mathit{InPhase}(D,P_1,\mathit{Ph}) \land \\ \mathit{YearsInProgram}(D,P_2,Y) \land \\ \end{array} \mathit{Rules}(38) \right) \\
\mathit{AdvisedBy}(P_1,P_2) \leftarrow \left( \begin{array}{l} \mathit{InPhase}(D,P_1,\mathit{Ph}) \land \\ \mathit{HasPosition}(D,P_2,\mathit{Po}) \land \\ \end{array} \mathit{Rules}(39) \right) \\
\mathit{AdvisedBy}(P_1,P_2) \leftarrow \left( \begin{array}{l} \mathit{InPhase}(D,P_1,\mathit{Ph}) \land \\ \mathit{InPhase}(D,P_2,\mathit{Ph}) \land \\ \end{array} \mathit{Rules}(40) \right) \\
\mathit{AdvisedBy}(P_1,P_2) \leftarrow \left( \begin{array}{l} \mathit{InPhase}(D,P_1,\mathit{Ph}) \land \\ \mathit{ProjectMember}(D,\mathit{Pr},P_2) \land \\ \end{array} \mathit{Rules}(41) \right) \\
\mathit{AdvisedBy}(P_1,P_2) \leftarrow \left( \begin{array}{l} \mathit{ProjectMember}(D,\mathit{Pr},P_1) \land \\ \mathit{Ta}(D,C,P_2,\mathit{Te}) \land \\ \end{array} \mathit{Rules}(42) \right) \\
\mathit{AdvisedBy}(P_1,P_2) \leftarrow \left( \begin{array}{l} \mathit{ProjectMember}(D,\mathit{Pr},P_1) \land \\ \mathit{TaughtBy}(D,C,P_2,\mathit{Te}) \land \\ \end{array} \mathit{Rules}(43) \right) \\
\mathit{AdvisedBy}(P_1,P_2) \leftarrow \left( \begin{array}{l} \mathit{ProjectMember}(D,\mathit{Pr},P_1) \land \\ \mathit{Publication}(D,\mathit{Ti},P_2) \land \\ \end{array} \mathit{Rules}(44) \right) \\
\mathit{AdvisedBy}(P_1,P_2) \leftarrow \left( \begin{array}{l} \mathit{ProjectMember}(D,\mathit{Pr},P_1) \land \\ \mathit{YearsInProgram}(D,P_2,Y) \land \\ \end{array} \mathit{Rules}(45) \right) \\
\end{array}
\]
\[
\begin{array}{l}
\mathit{AdvisedBy}(P_1,P_2) \leftarrow \left( \begin{array}{l} \mathit{ProjectMember}(D,\mathit{Pr},P_1) \land \\ \mathit{HasPosition}(D,P_2,\mathit{Po}) \land \\ \end{array} \mathit{Rules}(46) \right) \\
\mathit{AdvisedBy}(P_1,P_2) \leftarrow \left( \begin{array}{l} \mathit{ProjectMember}(D,\mathit{Pr},P_1) \land \\ \mathit{InPhase}(D,P_2,\mathit{Ph}) \land \\ \end{array} \mathit{Rules}(47) \right) \\
\mathit{AdvisedBy}(P_1,P_2) \leftarrow \left( \begin{array}{l} \mathit{ProjectMember}(D,\mathit{Pr},P_1) \land \\ \mathit{ProjectMember}(D,\mathit{Pr},P_2) \land \\ \end{array} \mathit{Rules}(48) \right) \\
\end{array}
\]
Regarding the variables, $D$ is a department name, $P_1$ and $P_2$ are persons, $C$ is
a course, {\em Po} is a position, {\em Te} is a term, {\em Ph} is a phase, {\em Pr}
is a project, {\em Ti} is a title, and {\em Y} is a year.

Finally, positive labels (label probability $1.0$) are from the real
instances of {\em AdvisedBy}. Negative labels (label probability $0.0$) are uniformly
drawn person-person pairs not present in {\em AdvisedBy}.

\section{Quality Task: Fact Extraction}
We set $\mathcal{T} := \mathit{Occurrence} \cup \mathit{Begin} \cup \mathit{During}
\cup \mathit{End}$ and 
$\mathcal{T}_l := \mathit{Begin} \cup \mathit{During} \cup \mathit{End}$. Also,
{\em Occurrence} is certain, whereas the other three relations are uncertain.
There are three types of rules.
First, for reconciling facts we have
\[
\begin{array}{l}
\mathit{IsMarriedToBegin}(E_1,E_2,T_1,T_2) \\
~~~\leftarrow \mathit{Begin}(\mathit{Pid}) \land \mathit{Occurrence}(\mathit{Pid},E_1,E_1,pp,T_1,T_2) \\
\mathit{IsMarriedToDuring}(E_1,E_2,T_1,T_2) \\
~~~\leftarrow \mathit{During}(\mathit{Pid}) \land \mathit{Occurrence}(\mathit{Pid},E_1,E_1,pp,T_1,T_2) \\
\mathit{IsMarriedToEnd}(E_1,E_2,T_1,T_2) \\
~~~\leftarrow \mathit{End}(\mathit{Pid}) \land \mathit{Occurrence}(\mathit{Pid},E_1,E_1,pp,T_1,T_2) \\
\mathit{WorksForClubBegin}(E_1,E_2,T_1,T_2) \\
~~~\leftarrow \mathit{Begin}(\mathit{Pid}) \land \mathit{Occurrence}(\mathit{Pid},E_1,E_1,pc,T_1,T_2) \\
\mathit{WorksForClubDuring}(E_1,E_2,T_1,T_2) \\
~~~\leftarrow \mathit{During}(\mathit{Pid}) \land \mathit{Occurrence}(\mathit{Pid},E_1,E_1,pc,T_1,T_2) \\
\mathit{WorksForClubEnd}(E_1,E_2,T_1,T_2) \\
~~~\leftarrow \mathit{End}(\mathit{Pid}) \land \mathit{Occurrence}(\mathit{Pid},E_1,E_1,pc,T_1,T_2) \\
\end{array}
\]
where $E_1$ and $E_2$ are entities, $T_1$ and $T_2$ are integers representing time
interval limits, {\em PId} is the pattern id, and {\em pp} and {\em pc} are constants
standing for the type pairs person-person and person-club, respectively.
The next rules enforce mutual exclusion
\[
\begin{array}{rcl}
\mathit{Constraint1}(\mathit{Pid}) & \leftarrow & \mathit{Begin}(\mathit{Pid})\land \mathit{During}(\mathit{Pid}) \\
\mathit{Constraint2}(\mathit{Pid}) & \leftarrow & \mathit{Begin}(\mathit{Pid})\land \mathit{End}(\mathit{Pid}) \\
\mathit{Constraint3}(\mathit{Pid}) & \leftarrow & \mathit{During}(\mathit{Pid})\land \mathit{End}(\mathit{Pid}) \\
\end{array}
\]
by labeling their resulting lineage with probability $0.0$.
Finally, we encode temporal precedence constraints by the rules
\[
\begin{array}{l}
\mathit{Constraint4}(E_1,E_2)
\leftarrow \left(
\begin{array}{l}
\mathit{IsMarriedToBegin}(E_1,E_2,T_1,T_2) \land \\
\mathit{IsMarriedToDuring}(E_1,E_2,T_3,T_4) \land \\
T_3 < T_2 \\
\end{array} \right) \\
\mathit{Constraint5}(E_1,E_2)
\leftarrow \left(
\begin{array}{l}
\mathit{IsMarriedToBegin}(E_1,E_2,T_1,T_2) \land \\
\mathit{IsMarriedToEnd}(E_1,E_2,T_3,T_4) \land \\
T_3 < T_2 \\
\end{array} \right) \\
\mathit{Constraint6}(E_1,E_2)
\leftarrow \left(
\begin{array}{l}
\mathit{IsMarriedToDuring}(E_1,E_2,T_1,T_2) \land \\
\mathit{IsMarriedToEnd}(E_1,E_2,T_3,T_4) \land \\
T_3 < T_2 \\
\end{array} \right) \\
\mathit{Constraint7}(E_1,E_2)
\leftarrow \left(
\begin{array}{l}
\mathit{WorksForClubBegin}(E_1,E_2,T_1,T_2) \land \\
\mathit{WorksForClubDuring}(E_1,E_2,T_3,T_4) \land \\
T_3 < T_2 \\
\end{array} \right) \\
\mathit{Constraint8}(E_1,E_2)
\leftarrow \left(
\begin{array}{l}
\mathit{WorksForClubBegin}(E_1,E_2,T_1,T_2) \land \\
\mathit{WorksForClubEnd}(E_1,E_2,T_3,T_4) \land \\
T_3 < T_2 \\
\end{array} \right) \\
\mathit{Constraint9}(E_1,E_2)
\leftarrow \left(
\begin{array}{l}
\mathit{WorksForClubDuring}(E_1,E_2,T_1,T_2) \land \\
\mathit{WorksForClubEnd}(E_1,E_2,T_3,T_4) \land \\
T_3 < T_2 \\
\end{array} \right) \\
\end{array}
\]
whose resulting lineage we label by probability $0.0$ as well.
Additionally, we use the 266 labels for textual patterns
and the 341 labels for facts from the original work.

\section{Runtime Task: SRL Methods}
\label{sec}
We synthetically set $\mathcal{T} = \mathcal{T}_l := \{t_0, \dots, t_{99}\}$
which are all uncertain tuples.
Then, we create a growing number of synthetic rule pairs following the pattern
\[
\begin{array}{l}
\mathit{Head}(c) \leftarrow t_i \land \neg t_j \land \neg t_k \\
\mathit{Head}(c) \leftarrow t_l \land \neg t_m \land \neg t_n \\
\end{array}
\]
such that $c$ is a constant indicating the rule id, and
$i,j,k,l,m,n$ are uniformly drawn random numbers from $0, \dots ,99$,
and displayed negations exist with probability $0.5$. Now, we uniformly draw
a synthetic probability label from $\{0.0, 1.0\}$ for $\mathit{Head}(c)$'s
lineage.

\section{Runtime Task: Gradient Methods}
We set $\mathcal{T}$ to be all of YAGO2's relations
and $\mathcal{T}_l := \mathit{LivesIn}$. All relations are uncertain, where
tuples in $\mathcal{T} \backslash \mathcal{T}_l$ have synthetic
probability values uniformly drawn from $[0,1]$. The only rule we have is
\[
\mathit{ToLabel}(P) \leftarrow \mathit{LivesIn}(P,L)
\]
whose resulting lineage formulas we label by synthetic probabilities
uniformly drawn from $[0,1]$.

\section{Runtime Task: Large Scale}
We set $\mathcal{T}$ to be all of YAGO2's relations. All relations are uncertain, where
tuples in $\mathcal{T} \backslash \mathcal{T}_l$ have synthetic
probability values uniformly drawn from $[0,1]$.
\subsection{P1}
Here, $\mathcal{T}_l := \mathit{ActedIn} \cup \mathit{WasBornIn}$ and the rules
are
\[
\begin{array}{l}
\mathit{Movie}(M) \leftarrow \mathit{ActedIn}(P,M) \\
\mathit{Creator}(P) \leftarrow \mathit{ActedIn}(P',M) \land \mathit{Created}(P,M) \\
\mathit{Location}(L) \leftarrow \mathit{WasBornIn}(P,L) \\
\mathit{Person}(P) \leftarrow \mathit{WasBornIn}(P,L) \\
\mathit{Person2}(P) \leftarrow \mathit{WasBornIn}(P,L) \land \mathit{IsLocatedIn}(L,L') \\
\end{array}
\]
whose resulting lineage formulas have synthetic probability labels
uniformly drawn from $[0,1]$.
\subsection{P2}
Here, $\mathcal{T}_l := \mathit{ActedIn} \cup \mathit{WasBornIn}$ and the rules
are
\[
\begin{array}{l}
\mathit{Movie}(M) \leftarrow \mathit{ActedIn}(P,M) \\
\mathit{Actor}(P) \leftarrow \mathit{ActedIn}(P,M) \land \mathit{Created}(P',M) \\
\mathit{Location}(L) \leftarrow \mathit{WasBornIn}(P,L) \\
\mathit{Person}(P) \leftarrow \mathit{WasBornIn}(P,L) \land \mathit{LivesIn}(P',L) \\
\end{array}
\]
whose resulting lineage formulas have synthetic probability labels
uniformly drawn from $[0,1]$.

\subsection{P3}
Here, $\mathcal{T}_l := \mathit{IsLocatedIn\_Transitive}$ and the only
rule is
\[
\mathit{Location}(L) \leftarrow \mathit{IsLocatedIn\_Transitive}(L,L')
\]
whose resulting lineage formulas have synthetic probability labels
uniformly drawn from $[0,1]$.

\section{Runtime Task: Objectives}
The setup is exactly the same as in Section~\ref{sec}.

\section{Additional Experiments}
In this section we present two additional experiments not present in the paper.
First, we vary the error rates $\epsilon_\mathit{rel}$ and 
$\epsilon_\mathit{abs}$ to investigate their impact on both runtime and quality
(Section~\ref{sec:error}).
Second, we run Algorithm 1 repeatedly to analyze its ability to find global optima
(Section~\ref{sec:global}).
As an overview, characteristics of the used learning problem instances are
available in the following table:
\begin{small}
\[
		\begin{array}{c|c|c|c|c|c|c|c|c}
		\text{Section} & \text{Figure} & \text{Source}
		  & |\mathcal{T}| & |\mathcal{T}_l| & |\mathcal{L}| & \text{Avg.}~T(\phi)
		  & l_i~\text{Boolean} & \text{Inconsistent} \\
		\hline
		  \ref{sec:error} & \ref{fig:epsilon}  & \text{UW-CSE} & 2,161 & 49 & 339 & 6.0 & \text{yes} & \text{yes}\\
		\hline
		   \ref{sec:global}& \ref{fig:errorConsistent}  & \text{UW-CSE} & 2,161 & 49 & 113 & 8.5 & \text{yes} & \text{no}\\
		\hline
		  \ref{sec:global} & \ref{fig:errorInconsistent}  & \text{UW-CSE} & 2,161 & 49 & 339 & 6.0 & \text{yes} & \text{yes}\\
		\end{array}
\]
\end{small}

\subsection{Varying Error Rates}
\label{sec:error}

\begin{figure}
	\centering sfsf
  	\includegraphics[width=0.48\textwidth]{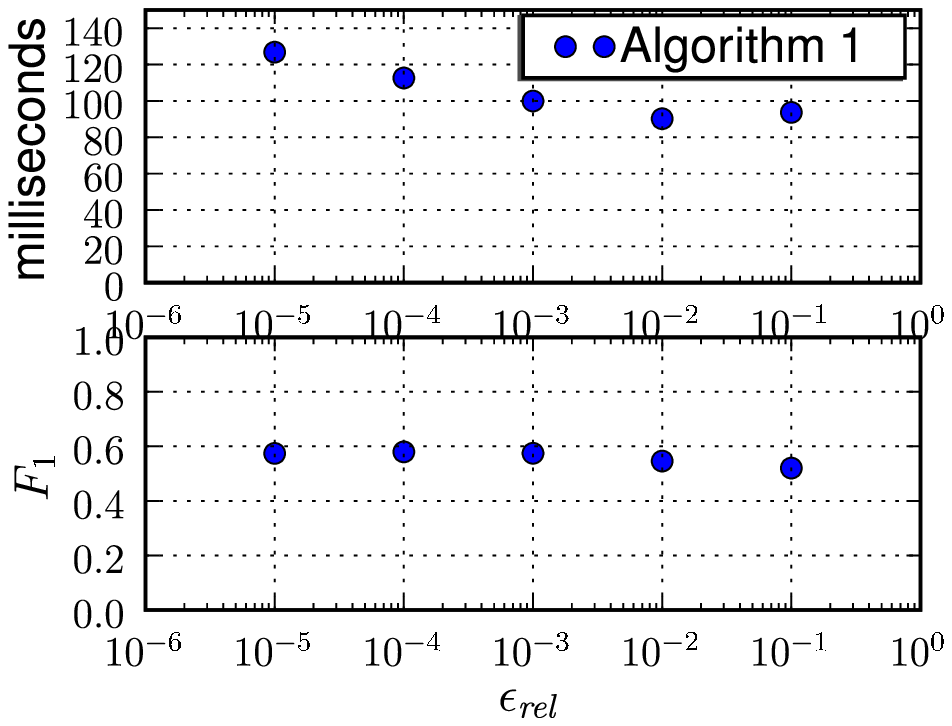}
	\caption{Varying Error Rates}
	\label{fig:epsilon}
\end{figure}

\noindent \textbf{Setup.}
We use Section~\ref{sec:uwcse}'s setup with twice as many negative labels
as positive ones, where we vary $\epsilon_\mathit{rel}$ from $10^{-1}$
to $10^{-5}$. Moreover, we set
$\epsilon_\mathit{abs} := \frac{\epsilon_\mathit{rel}}{100}$.
In comparison, all other experiments had fixed $\epsilon_\mathit{rel} = 10^{-4}$ and
$\epsilon_\mathit{abs} = 10^{-6}$.

\noindent \textbf{Results.}
In Figure~\ref{fig:epsilon} we display both runtime and
quality in terms of the $F_1$ measure.

\noindent \textbf{Discussion.}
While inspecting Figure~\ref{fig:epsilon} we realize that decreasing the
error-rates the runtime increases slightly. Furthermore, $F_1$ is worse
for $\epsilon_\mathit{rel} > 10^{-3}$. But, if we compare the overall runtime
presented in the paper's Figure 4(c), we realize that Algorithm 1 consumes
only small portions of the total runtime of about 3 seconds.

\subsection{Finding Global Optima}
\label{sec:global}
\begin{figure*}
	\subfigure[Consistent Instance]{ \label{fig:errorConsistent}
  	\includegraphics[width=0.48\textwidth]{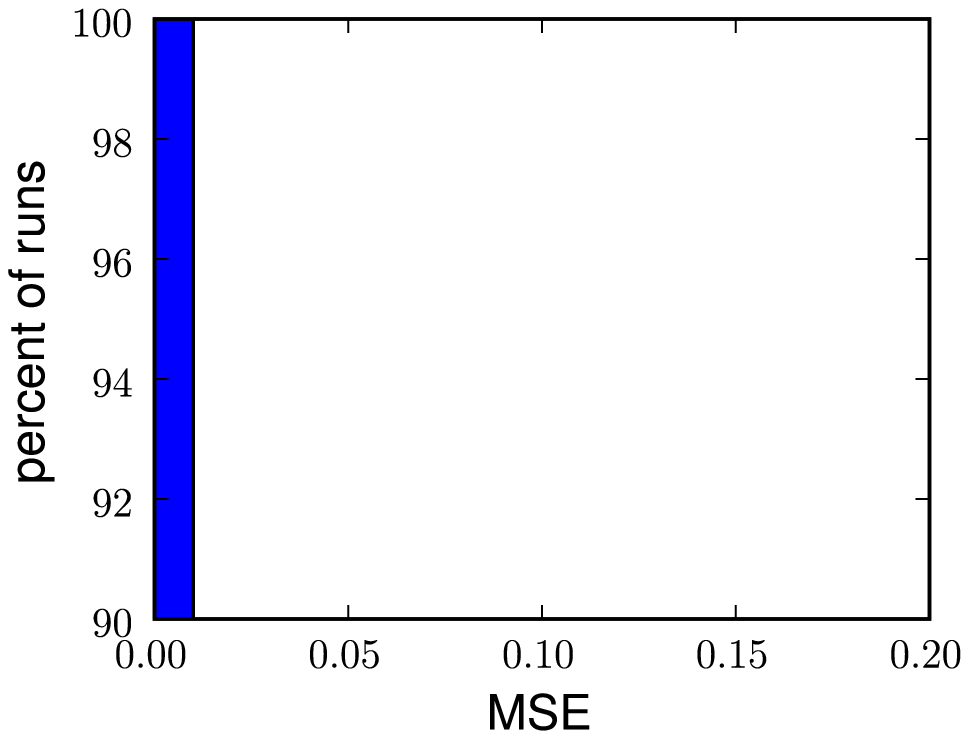}
	}
	\subfigure[Inconsistent Instance]{ \label{fig:errorInconsistent}
  	\includegraphics[width=0.48\textwidth]{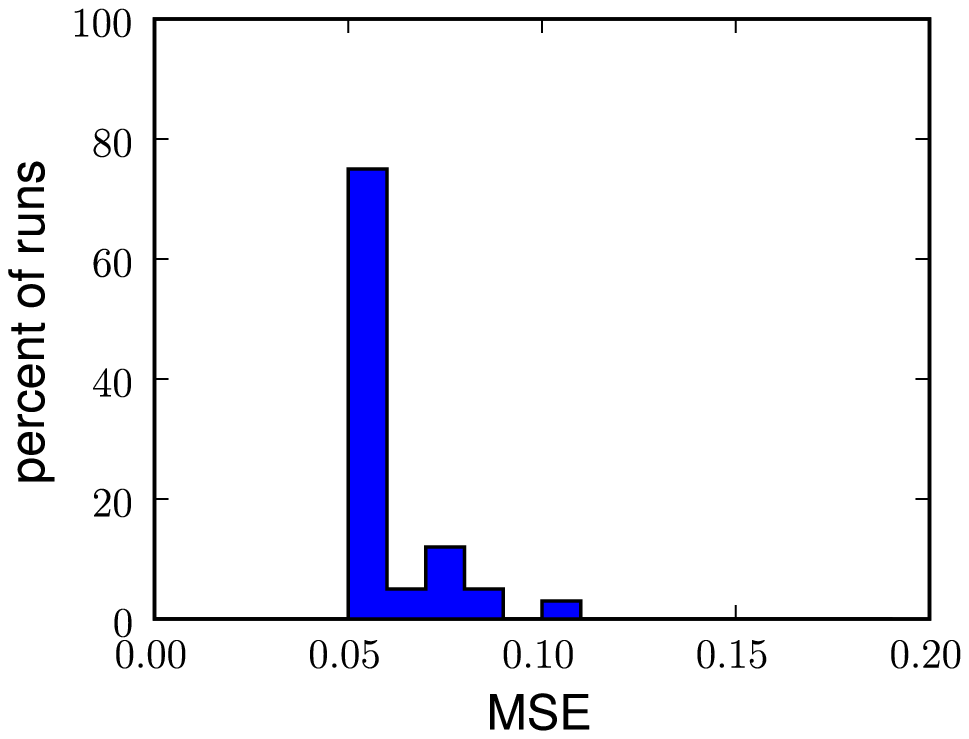}
	}
	\caption{Finding Global Optima}
	\label{fig:global}
\end{figure*}

\noindent \textbf{Setup.}
We use Section~\ref{sec:uwcse}'s setup with two different instances, a consistent
one (only positive labels) and an inconsistent one (positive and negative labels).

\noindent \textbf{Results.}
In Figure~\ref{fig:global} we depict histograms of the resulting
mean-squared-error (MSE) of 100 runs on each instance.

\noindent \textbf{Discussion.}
Algorithm 1 always finds solutions extremely close to the global optimum ($0.0$) in the consistent instance (Figure~\ref{fig:errorConsistent}).
The inconsistent instance, however, has global optima of values larger than $0.0$.
In this case Algorithm 1 converges very close to a (probably) global optimum
in $78\%$ of the runs.

%% file: intro.tex
\section{Introduction}
The increasing availability of large, uncertain datasets, for example arising from
imprecise sensor readings, information extraction or data integration
applications, has led to a recent advent in the research on probabilistic
databases (PDBs)~\cite{PDBbook}. PDBs adopt scalable techniques for processing
queries expressed in SQL, Relational Algebra or Datalog from their
deterministic counterparts. However, already for fairly simple
select-project-join (SPJ) queries, computing a query answer's confidence (in
the form of a mar\-ginal probability) remains a $\#\mathcal{P}$-hard
problem~\cite{DalviDichotomyConjQuerSharpP,Dalvi:2007:EQE:1285882.1285906}.
%also in PDBs
Consequently, the majority of scientific works in this area is centered around
the problem of confidence computation, either by investigating tractable
subclasses of query plans~\cite{DalviDichotomyConjQuerSharpP,Dalvi:2007:EQE:1285882.1285906,PDBbook},
various knowledge compilation techniques~\cite{JhaCompiling},
or general approximation techniques~\cite{OlteanuHK10}. 
Moreover, nearly all the works we are aware of (except~\cite{DerivingPDBs}) assume the
probability values of base tuples stored in the PDB to be given.
Although stated as a major challenge already in~\cite{PDBdiamondsDirt} by Dalvi, R\'e and Suciu, 
to this date,
incorporating user feedback in order to create, update or
clean a PDB has been left as future work.
%To this date,
%learning or altering the tuple probabilities in order to create, update or
%clean a PDB has been left as future work. In~\cite{PDBdiamondsDirt}, for
%example, Dalvi, R\'e and Suciu explicitly motivate for the inclusion of user
%feedback in order to ``clean'' the underlying probabilistic database. %This is
% a difficult problem, which to date has not yet been solved."

\noindent\textbf{Problem Setting.} In this work, we address the problem of updating or cleaning a PDB by learning tuple probabilities from %user- or application-provided 
labeled lineage formulas. This problem can be seen as the inverse problem to confidence computations in PDBs: given a set of Boolean lineage formulas, each labeled with a probability, learn the probability values associated with the base tuples in this PDB, such that the marginal probabilities of the lineage formulas again yield their probability labels. The labels serving as input can equally result from human feedback, an application running on top of the PDB, or they can be obtained from a provided set of consistency constraints.

\noindent\textbf{Related Techniques.} 
Our work is closely related to learning the parameters (i.e., weights) of
probabilistic-relational models in the field of Statistical Relational Learning
(SRL) \cite{SRLintroBook}. SRL comes with a plethora of individual approaches
(most notably Markov Logic~\cite{WeightLearningMLN,MLN,Singla:2005:DTM:1619410.1619472} and
%the probabilistic logic programming engine
ProbLog~\cite{ProbLog}), but due to the
generality of these techniques, which are designed to support large fragments
of first-order logic, it is difficult to scale these to database-like
instance sizes.
%instances with millions of tuples and hundreds of thousands of labeled training
%data.
In this work, we focus on relational (and probabilistic) data as input
and on the core operations expressible in Relational Algebra or Datalog for
query processing. 
Moreover, we show that our approach subsumes previously raised problems of deriving PDBs from incomplete databases \cite{DerivingPDBs} as well as of enforcing constraints over PDBs via conditioning the base tuples onto a given set of consistency constraints \cite{KochConditioningPDB}.
We illustrate our setting by the following running example.
\begin{example} \label{ex:intro}
Our running example resembles a simple in\-for\-mation-extraction setting, in which we employ a set of textual patterns to extract facts from various Web domains. The references to the involved patterns and the domains, as well as the extracted facts, together form the PDB shown in Figure~\ref{fig:exampleDB}.
%{\em Spielberg}, as it is . % depicts a probabilistic database describing how facts about him were extracted from webpages.
\begin{figure}[h]
	\[
	\begin{array}{c|c|c|c|c|c}
	\multicolumn{1}{c}{} & \multicolumn{4}{c}{\text{WonPrizeExtraction}} & \\
	\cline{2-5}
	& \text{Subject} & \text{Object} & \text{Pid} & \text{Did} &  p\\
	\cline{2-5}
	t_1 & \mathit{Spielberg} & \mathit{AcademyAward} & 1 & 1 & 0.6\\
	t_2 & \mathit{Spielberg} & \mathit{AcademyAward} & 2 & 1 & 0.3\\
	\cline{2-5}
	\end{array}
	\]
	\vspace{-3mm}
	\[
	\begin{array}{c|c|c|c|c|c}
	\multicolumn{1}{c}{} & \multicolumn{4}{c}{\text{BornInExtraction}} & \\
	\cline{2-5}
	& \text{Subject} & \text{Object} & \text{Pid} & \text{Did} &  p\\
	\cline{2-5}
	t_3 & \mathit{Spielberg} & \mathit{Cinncinati} & 3 & 1 & 0.7\\
	t_4 & \mathit{Spielberg} & \mathit{LosAngeles} & 3 & 2 & 0.4\\
	\cline{2-5}
	\end{array}
	\]
	\vspace{-3mm}
	\[
	\begin{array}{c|c|c|ccc|c|c|c}
	\multicolumn{1}{c}{} & \multicolumn{2}{c}{\text{UsingPattern}} & & & \multicolumn{1}{c}{} & \multicolumn{2}{c}{\text{FromDomain}} & \\
	\cline{2-3} \cline{7-8}
	& \text{Pid} & \text{Pattern} & p & & & \text{Did} & \text{Domain} & p\\
	\cline{2-3} \cline{7-8}
	t_5 & 1 & \mathit{Received} & ? & & t_8 & 1 & \mathit{Wikipedia.org} & ?\\
	t_6 & 2 & \mathit{Won} & ? & & t_9 & 2 & \mathit{Imdb.com} & ?\\
 	\cline{7-8}
	t_7 & 3 & \mathit{Born} & ? & & \multicolumn{4}{c}{}\\
	\cline{2-3}
	\end{array}
	\]
	\vspace{-4mm}
	\caption{An Example Probabilistic Database}
	\label{fig:exampleDB}
\end{figure}
The fact captured by $t_1$, for example, expresses that {\em Spielberg} won an {\em AcademyAward}
with a given probability value of $0.6$, which we consider to be provided as input to our database.
In contrast, the probability values of tuples in {\em UsingPattern} and {\em FromDomain}
are unknown (as indicated by the question marks). We thus are unsure about the reliability---or trustworthiness---of the extraction patterns and the Web domains that led to the extraction of our remaining facts, respectively. 
%Learning these missing probabilities is the subject of this paper.

By joining the {\em WonPrizeExtraction} relation with
{\em UsingPattern} and {\em FromDomain} on {\em Pid} and {\em Did}, respectively,
we can see that $t_1$ was extracted from {\em Wikipedia.org} using the textual pattern {\em Received}.
We express this join via the following deduction rule (in Datalog-style syntax):
\begin{equation} \label{eq:deducePrize}
\mathit{WonPrize}(S,O) \leftarrow
\left(\!\!\!\begin{array}{c}
\mathit{WonPrizeExtraction}(S,O,\mathit{Pid},\mathit{Did}) \\
\land \mathit{UsingPattern}(\mathit{Pid},P) \\
\land \mathit{FromDomain}(\mathit{Did}, D)\\
\end{array}\!\!\!\right)
\end{equation}
Analogously, we reconcile fact extractions for the {\em BornIn} relation as follows:
\begin{equation} \label{eq:deduceBorn}
\mathit{BornIn}(S,O) \leftarrow
\left(\!\!\!\begin{array}{c}
\mathit{BornInExtraction}(S,O,\mathit{Pid},\mathit{Did}) \\
\land \mathit{UsingPattern}(\mathit{Pid},P) \\
\land \mathit{FromDomain}(\mathit{Did}, D)\\
\end{array}\!\!\!\right)
\end{equation}
Instantiating (i.e., ``grounding'') Rules~\eqref{eq:deducePrize}
and~\eqref{eq:deduceBorn}
against the base tuples of Figure~\ref{fig:exampleDB}
yields the new tuples {\em BornIn(Spiel\-berg, Cinncinati)},
{\em BornIn(Spielberg, LosAngeles)},
and {\em Won\-Prize(Spielberg,AcademyAward)}. 
Figure~\ref{fig:exampleLineage} shows these new tuples along with their Boolean lineage formulas, which
capture their logical dependencies with the base tuples. % and the query answers derived from the two rules.
%, how the new tuples were derived, i.e.\ disjunctions ($\lor$) represent alternatives and conjuctions ($\land$) joins.
\begin{figure}[h]
	\begin{center}
	\includegraphics[width=0.45\textwidth]{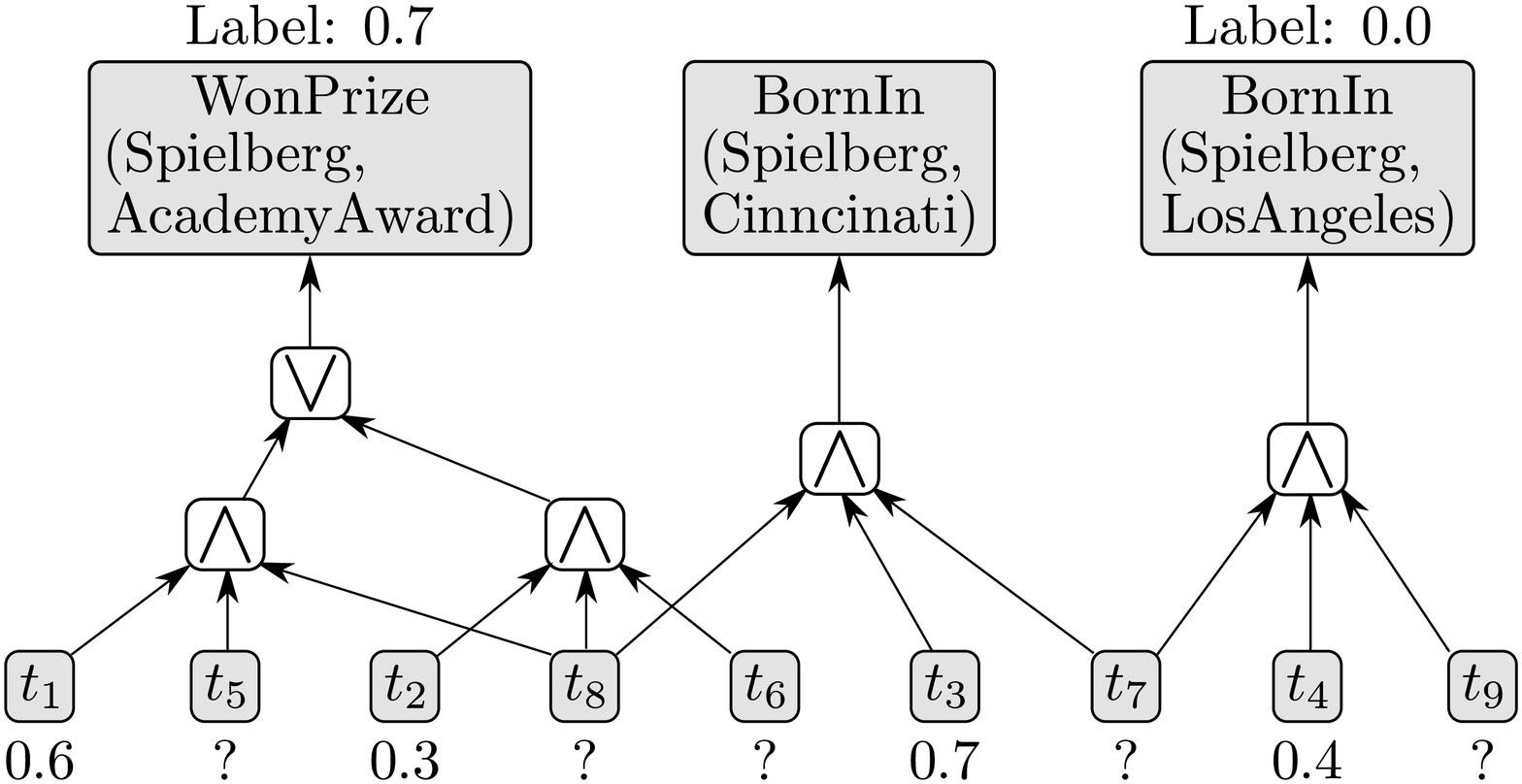}
	\end{center}
	\vspace{-6mm}
	\caption{Example Lineages and Labels}
	\label{fig:exampleLineage}
\end{figure}
A closer look at the new tuples reveals, however, that not all of them are correct.
For instance, {\em BornIn(Spielberg,LosAngeles)}
is wrong, so we label it with the probability of $0.0$. Moreover,
{\em WonPrize(Spielberg,AcademyAward)} is likely
correct, hence we label it with the probability of $0.7$, as shown
on top of Figure~\ref{fig:exampleLineage}.
Given the probability labels of the query answers, the goal of the learning procedure is to learn the base tuples' unknown probability
values for {\em UsingPattern} and {\em FromDomain} (while leaving the ones for {\em WonPrizeExtraction} and {\em BornInExtraction} unchanged),
such that the lineage formulas again produce the given probability labels.
\end{example}

%We summarize the contributions of this paper as follows.
\vspace{4mm}
\noindent \textbf{Contributions.}
We summarize the contributions of this work as follows.
\squishlist
\item To our knowledge, we present the first approach to tackle the problem of
 	 learning unknown (or missing) tuple probabilities from labeled lineage formulas in the context of PDBs.
	 In Section~\ref{sec:weightLearning}, we formally define the learning problem
   and analyze its properties from a theoretical perspective.
   %We derive insights on the number of required labels, and
	 %the impact of a labeled lineage formula's structure on the problem.
\item We formulate the learning problem as an optimization problem, devise
	two different objective functions for solving it, and discuss both in Section~\ref{sec:solving}.
\item In Section~\ref{sec:algorithm}, we present a learning algorithm based on stochastic gradient descent, which scales to problem instances
 	 with hundreds of thousands of labels and millions of tuples to learn
	 the probability values for (Section~\ref{sec:exp:scalability}).
\item In Section~\ref{sec:relatedProblems}, we show that the learning
   problem supports prior probabilities of base tuples which can be incorporated
	 to update and clean PDBs. Also, we demonstrate that the learning problem
	 subsumes both learning from incomplete databases~\cite{DerivingPDBs} and applying
	 constraints to PDBs~\cite{KochConditioningPDB}.
 \item Additionally, we perform an
   experimental evaluation on three different real world datasets 
	 as well as on synthetic data (see Figure~\ref{table:datasets}), where we
	 compare our approach to various techniques based on SRL, reasoning in
	 information extraction, and
	 optimization (Section~\ref{sec:experiments}).
% \item To the best of our knowledge, no publicly available dataset for probabilistic
%   databases with learned tuple probabilities exists. We publish the source
%	 code of the described system\footnote{\label{footnote:sourceCode}\href{http://www.mpi-inf.mpg.de/~mdylla/}{http://www.mpi-inf.mpg.de/\textasciitilde mdylla/}}, hence enabling the
%	 creation of such datasets by the scientific community.
\squishend

%% file: relatedWork.tex
\section{Related Work}
In the following, we briefly review a number of related works from the areas of SRL and PDBs, which we believe are closest to our work. %Learning the parameters of statistical models %traditionally , focus on the
%database-oriented view where this problem has been largely
%ignored.
%A large body of literature in machine learning, with its subfield
%statistical relational learning~\cite{SRLintroBook} being closest to
%PBDs, considers learning. 

\noindent\textbf{Machine Learning.}
Many machine learning approaches have been applied to large scale data sets (see~\cite{ScalingUpML} for an overview). However, the scalable methods tend to not offer a declarative language (similar to deduction rules or constraints) in order to induce correlations among facts, as queries and lineage do in PDBs. In contrast, in the subfield of SRL~\cite{SRLintroBook}, correlations between ground atoms (similar to base tuples in PDBs) are often
induced by logical formulas (similar to lineage in PDBs).
But in turn, these methods lack scalability.
{\em Markov Logic Networks} (MLNs)~\cite{MLN} (and their
learning techniques~\cite{WeightLearningMLN,Singla:2005:DTM:1619410.1619472})
are built on an open-world assumption, which
%By following the default interpretation
%of formulas in first-order logic, the set of ground atoms is obtained by
%replacing the variables inside a literal by all possible combinations of
%constants that the type signature of the respective predicate admits. This
%often results in a blow-up of the grounded network structure that is
instantiates all combinations of constants, often resulting in a blow-up
incompatible to data\-base-like instance sizes. 
Even a very efficient implementation of MLNs, {\em Markov: TheBeast}~\cite{riedel08},
does not meet the scalability required for databases (see Section~\ref{sec:exp:scalingVSsrl}).
As opposed to MLNs, {\em ProbLog}~\cite{ProbLog} computes marginal probabilities while relying
on SLD proofs, which makes it very similar to PDBs with their
closed-world assumption and deductive grounding techniques.
However, also its learning procedure~\cite{ProbLogLearning} does not scale well to large datasets (see Section~\ref{sec:exp:scalingVSsrl}).
Within the ProbLog framework, \cite{Gutmann:2008:PLP:1431932.1431985} proposes the most similar
approach to ours, however lacking both a theoretical analysis and large scale experiments.

\noindent\textbf{Probabilistic Databases.}
A number of PDB engines, including {\em MystiQ}~\cite{DBLP:conf/sigmod/BoulosDMMRS05},
{\em MayBMS}~\cite{antova2008}, %and SPROUT~\cite{OlteanuHK09},
and {\em Trio}~\cite{DBLP:journals/vldb/BenjellounSHTW08} %, and PrDB~\cite{Kanagal:2010,SenDG09},
have been released as open-source prototypes in recent years and found a wide
recognition in the database community. Due to the hardness of computing probabilities 
for query answers, a main focus of these approaches lies
%---besides the system
%aspect of supporting probabilistic inference in a database-like
%environment---
in finding tractable subclasses of query
plans~\cite{DalviDichotomyConjQuerSharpP,Dalvi:2007:EQE:1285882.1285906,PDBbook}
for which probability computations can be done in polynomial time. 
%A recent trend towards scalable inference lies in the adoption of knowledge
%compilation techniques from the field of SRL, by compiling Boolean formulas
%into more succinct representation formalisms such as OBDDs, FBDDs, and even
%first-order d-DNNFs~\cite{JhaCompiling}.
A recent trend towards scalable inference is compiling Boolean formulas into
more succinct representation formalisms such as OBDDs~\cite{JhaCompiling}.
\cite{QueryEvalIncidenceAlgebra,JhaCompiling}, for example, develop an entire
lattice of algebras and compilation techniques over unions of conjunctive
queries (UCQs) which admit for polynomial-time inference. 
MarkoViews~\cite{markoViews} represent another step towards SRL, by introducing uncertain views, where probabilities depend on the input tuples, but---still---do not tackle the actual learning problem. Also, \cite{WangFGH10} circumvent the learning problem by enabling direct querying of Conditional Random Fields via a probabilistic database.
%approximation~\cite{OlteanuHK10}, and the theoretically general framework~\cite{QueryEvalIncidenceAlgebra}.

\noindent\textbf{Creating Probabilistic Databases.}
There are very few works on the actual creation of PDBs. The authors of \cite{DerivingPDBs} induce a probabilistic database by estimating probabilities from a given incomplete database, a problem that is subsumed by our definition of the learning problem (see Section~\ref{sec:incompleteAndLearning}). Enforcing consistency constraints by conditioning the base tuples of a PDB~\cite{KochConditioningPDB} onto these constraints allows for altering the tuple probabilities. Since conditioning lacks support for non-Boolean or inconsistent constraints,
we can show that our work also subsumes this problem, but not the other way round (see Section~\ref{sec:constraintsAndLearning}). Similarly, incorporating user feedback by means of probabilistic data integration, as in~\cite{DataIntegrationUserFeedback}, focuses on consistent, Boolean-only labels. 
%We believe that our approach has a strong application to data cleaning (see~\cite{DBLP:series/synthesis/2012Fan} for an overview) in deterministic databases, where the goal is to identify (and hence to create) a consistent database from inconsistent input data.

\noindent\textbf{Lineage \& Polynomials.}
The theoretical analysis of our learning problem (Section~\ref{sec:weightLearning}) is based on computing marginal probabilities via polynomials.
%The approach presented in~\cite{ProvenanceSemirings} exhibits similarities to our encoding, but focuses on representing operations expressible in Relational Algebra as semiring operations over polynomials. \cite{ProvenanceSemirings} does not investigate an application of this encoding for probabilistic inference.
Similarly, the authors of~\cite{ProvenanceSemirings} used semirings over polynomials to
model provenance, where lineage is a special case.
Also, {\em Sum-Product Networks}~\cite{sumProduct} investigates tractable graphical models by representing these as polynomial expressions with polynomially many terms.
%\cite{Michelakis:2009:UMR:1559845.1559858} bayesian formulation of weight learning for extraction rules, lineage?

%% file: model.tex
\section{Probabilistic Databases}
\label{sec:ProbabilisticDatabases}
In this section, we introduce our data model which follows the common possible-worlds
semantics over tuple-indepen\-dent probabilistic databases with lineage~\cite{PDBbook},
which is closed and complete~\cite{DBLP:journals/vldb/BenjellounSHTW08}.
Throughout this section, we assume that the probabilities of
all base tuples are known and fixed (i.e., even for $t_5$--$t_9$ in Example~\ref{ex:intro}).
Later, in Section~\ref{sec:weightLearning} we relax this view
to address the learning problem.

\noindent \textbf{Probabilistic Database.}
We define a {\em tuple-independent probabilistic database}~\cite{PDBbook}
$(\mathcal{T}, p)$ as a pair consisting of 
a finite set of {\em base tuples} $\mathcal{T}$ and a {\em probability measure} $p:
\mathcal{T} \rightarrow [0,1]$, which assigns a {\em probability value} $p(t)$ to each uncertain
tuple $t \in \mathcal{T}$. As in a regular database, we assume the set of
tuples $\mathcal{T}$ to be partitioned into a set of
{\em extensional relations} (see, e.g., {\em WonPrizeExtraction}, {\em BornInExtraction},
{\em UsingPattern}, and {\em FromDomain} in Example~\ref{ex:intro}).
The probability value $p(t)$ of a base tuple $t$ thus denotes the confidence in the existence of the tuple
in the database, i.e., a higher value $p(t)$
denotes a higher confidence in $t$ being valid.

\noindent \textbf{Possible Worlds.}
Assuming independence among all %the Boolean random variables associated with the 
base tuples
$\mathcal{T}$, the {\em probability} $P(\mathcal{W}, \mathcal{T})$ of
a {\em possible world}
$\mathcal{W} \subseteq \mathcal{T}$ is defined as follows.
\begin{eqnarray} \label{eq:ProbPossibleWorld}
P(\mathcal{W}, \mathcal{T}) := \prod_{t \in \mathcal{W}} p(t) \prod_{t \in \mathcal{T} \backslash \mathcal{W}} (1-p(t))
\end{eqnarray}
In the absence of any constraints (compare to Subsection~\ref{sec:constraintsAndLearning}) that would restrict this set of possible worlds, any subset $\mathcal{W}$ of tuples in $\mathcal{T}$ forms a valid possible world (i.e., a possible
instance) of the probabilistic database. Hence, there are exponentially many possible worlds.

\noindent\textbf{Deduction Rules.}
To support query answering over a PDB, we employ deduction rules (see, e.g., Rules~\eqref{eq:deducePrize} and~\eqref{eq:deduceBorn}), which we express in a Datalog-style notation.
Syntactically, these deduction rules have the shape of a logical implication
with exactly one positive head literal and a conjunction of both
positive and negative literals in the body. Formally, the class of rules
we support corresponds to {\em safe}, {\em non-recursive} Datalog programs,
which also coincides with the core operations expressible in the Relational Algebra~\cite{abiteboul-book}.
\begin{definition}
A {\em deduction rule} is a logical rule of the form
\[
R(\bar{X}) \leftarrow \bigwedge_{i = 1,\dots,n} R_i(\bar{X}_i) \land \bigwedge_{j=1,\dots,m} \neg R_j(\bar{X}_j) 
~~\land \Phi(\bar{X}')
\]
where
\squishlist
	\item[1.]
		$R$ denotes the head literal's intensional relation, whereas $R_i$, $R_j$ may refer
		to both intensional or extensional relations;
	\item[2.] $n \geq 1$, $m \geq 0$, thus requiring at least one positive relational literal;
	\item[3.] $\bar{X}$, $\bar{X}_i$, $\bar{X}_j$, and $\bar{X}'$ denote tuples of
		variables and constants, where $\mathit{Var}(\bar{X}), \mathit{Var}(\bar{X}_j) , \mathit{Var}(\bar{X}') \subseteq \bigcup_i \mathit{Var}(\bar{X}_i)$;
	\item[4.] $\Phi(\bar{X}')$ is a conjunction of arithmetic predicates, such as `` $=$'', ``$\neq$'', and ``$<$''.
\squishend
\end{definition}

%\begin{example}
%Assuming an intensional relation {\em Died} we
%can deduce famous and alive people by writing
%\[
%\mathit{Famous}(S) \leftarrow \mathit{WonPrize}(S, O) \land \neg \mathit{Died}(S)
%\]
%which features intensional relations and a negation in the body.
%$\diamond$
%\end{example}

\noindent\textbf{Lineage.} %Theoretically, we can evaluate the deduction rules over each possible world, which however is too expensive. Instead,
We utilize {\em data lineage} to represent the
logical dependencies between base tuples in $\mathcal{T}$
and tuples derived from the deduction rules (see Figure~\ref{fig:exampleLineage}).
In analogy to~\cite{PDBbook}, we consider lineage as a Boolean
formula. It relates each derived tuple (or ``query answer'') with the base tuples $\mathcal{T}$
via the three Boolean
connectives $\land$, $\lor$ and $\neg$, which reflect the semantics of
the relational operations that were applied to derive that tuple.
Specifically, we employ \squishlist
\item a conjunction ($\land$) that connects the relational literals in the body of a deduction rule;
\item a negation ($\neg$) for a negated relational literal in the body of a deduction rule;
\item a disjunction ($\lor$) whenever the same tuple is obtained from the head literals of two or more deduction rules;
\item a Boolean (random) variable $t$ representing a tuple in $\mathcal{T}$ whenever
an extensional literal matches this tuple.
\squishend
For a formal definition of lineage
in combination with Datalog rules and relational operators, we
refer the reader to~\cite{dylla13} and~\cite{PDBbook}, respectively.

\begin{example} 
\label{ex:lineage}
In Figure~\ref{fig:exampleLineage}, the conjunctions ($\land$)
are obtained from instantiating the conjunctions in Rule~\eqref{eq:deducePrize} and \eqref{eq:deduceBorn}'s bodies.
Because two instances of Rule~\eqref{eq:deducePrize} result in
the same derived tuple {\em WonPrize(Spielberg,AcademyAward)},
the disjunction ($\lor$) connects the two instantiated bodies.
$\diamond$
\end{example}

\noindent \textbf{Marginal Probabilities.}
We say that a possible world $\mathcal{W}$ {\em entails} a Boolean lineage formula $\phi$, denoted as
$\mathcal{W} \models \phi$, if it represents a satisfying truth assignment to $\phi$
by setting all tuples
in $\mathcal{W}$ to {\em true} and all tuples in $\mathcal{T} \backslash \mathcal{W}$ to
{\em false}.
%For the following steps, we utilize logical entailment to define an {\em indicator function} $\mathbf{1}_{\phi}$ that takes the value $1$ if the
%possible world $\mathcal{W}$
%entails $\phi$, and $0$ otherwise:
%\[
%\mathbf{1}_{\phi}(\mathcal{W}) :=
%\left\{
%\begin{array}{cl}
%1 & \text{if}~\mathcal{W} \models \phi \\
%0 & \text{if}~\mathcal{W} \not\models \phi \\
%\end{array}
%\right.
%\]
Then, we can compute the {\em marginal probability} of any Boolean formula $\phi$ over
tuples in $\mathcal{T}$ as the sum of the probabilities of all the possible worlds
$\mathcal{W}\subseteq\mathcal{T}$ that entail $\phi$:
\begin{equation} \label{eq:lineageConf}
P(\phi) := \sum_{\mathcal{W}\subseteq \mathcal{T}, \mathcal{W} \models \phi} ~\underbrace{P(\mathcal{W}, \mathcal{T})}_{\text{via Eq.}~\eqref{eq:ProbPossibleWorld}}
\end{equation}
%Again, in the absence of any constraints that would restrict the set of possible worlds, the probability value $p(t)$ attached to a base tuple $t \in \mathcal{T}$ coincides with its marginal probability $P(t)$.

%The sum of Equation~\ref{eq:lineageConf} however ranges over exponentially many terms.
To avoid the exponential cost involved in following Equation~\eqref{eq:lineageConf}, we can---in many cases---compute the marginal probability $P(\phi)$
directly on the structure of the lineage formula $\phi$~\cite{PDBbook}. Let
$T(\phi) \subseteq \mathcal{T}$ denote the set of base tuples occurring in $\phi$.
\begin{equation} \label{eq:confComp}
\begin{small}
\!\!
\begin{array}{rcl|c}
\multicolumn{3}{c|}{\text{Definition}} & \text{Condition}\\
\hline
P(t) \!\!& \!\!:= \!\!&\!\! p(t) & t \in \mathcal{T} \\
P(\bigwedge_i \phi_i) \!\!&\!\! := \!\!&\!\! \prod_i P(\phi_i) & i\!\neq\! j \Rightarrow \mathit{T}(\phi_i)\! \cap\! \mathit{T}(\phi_j)\! =\! \emptyset \\
P(\bigvee_i \phi_i) \!\!&\!\! := \!\!&\!\! 1 - \prod_i (1-P(\phi_i)) & i\!\neq\! j \Rightarrow \mathit{T}(\phi_i)\! \cap\! \mathit{T}(\phi_j)\! =\! \emptyset \\
%P(\phi \land \psi) \!\!&\!\! := \!\!&\!\! 0 & \phi \land \psi \equiv \mathit{false}\\
P(\phi \lor \psi) \!\!&\!\! := \!\!&\!\! P(\phi) + P(\psi) & \phi \land \psi \equiv \mathit{false}\\
P(\neg \phi) \!\!&\!\! := \!\!&\!\! 1 - P(\phi) & \\
\end{array}
\end{small}
\end{equation}
The first line captures the case of a base tuple $t$,
for which we return its attached probability value $p(t)$.
The next two lines handle {\em independent-and} and {\em independent-or}
operations for conjunctions and disjunctions over variable-disjoint subformulas $\phi_i$,
respectively. In the following line, we
address disjunctions for two subformulas $\phi$ and $\psi$ that denote disjoint probabilistic events
(known as {\em disjoint-or}~\cite{PDBbook}).
The last line finally handles negation.
Equation~\eqref{eq:confComp}'s definition of $P(\phi)$ runs in linear
time in the size of $\phi$. However, for general Boolean formulas, computing $P(\phi)$ is
$\#\mathcal{P}$-hard~\cite{DalviDichotomyConjQuerSharpP,PDBbook}. This
becomes evident if we consider Equation~\eqref{eq:shannon},
called {\em Shannon expansion}, which is a form of variable elimination that is applicable to any Boolean formula:
\begin{equation}\label{eq:shannon}
P(\phi) := p(t) \cdot P(\phi_{[t \rightarrow \mathit{true}]}) + (1 - p(t))
\cdot P(\phi_{[t \rightarrow \mathit{false}]})
\end{equation}
Here, the notation $\phi_{[t \rightarrow \mathit{true}]}$ for a tuple
$t \in T(\phi)$ denotes that we replace all occurrences of $t$ in $\phi$ by {\em true}
(and {\em false}, respectively). Repeated
applications of Shannon expansions may however result in an exponential
increase of $\phi$. The hardness of computing $P(\phi)$ for general
propositional formulas has been addressed by various techniques~\cite{PDBbook}, such as
knowledge compilation~\cite{JhaCompiling} %, top-$k$~\cite{dylla13}
or approximation~\cite{OlteanuHK10}.

\begin{example}
Consider $P( (t_1 \land t_5 \land t_8) \lor (t_2 \land t_6 \land t_8) )$
and assume $p(t_5) = 0.5$, $p(t_6) = 0.6$, and $p(t_8) = 0.8$
in addition to the known tuple probabilities shown in Figure~\ref{fig:exampleDB}. First,
Line 3 of Equation~\eqref{eq:confComp} is not applicable, since $t_8$ occurs on both sides.
So we apply a Shannon expansion, yielding
$p(t_8) \cdot P( (t_1 \land t_5) \lor (t_2 \land t_6) ) + (1 - p(t_8)) \cdot P(\mathit{false})
= 0.8 \cdot P( (t_1 \land t_5) \lor (t_2 \land t_6) )$, where we used $P(\mathit{false}) = 0.0$.
Next, we apply Line 3 of Equation~\eqref{eq:confComp} which results in
$0.8 \cdot (1 - (1 - P(t_1 \land t_5)) \cdot (1 - P(t_2 \land t_6) ))$.
Then, two applications of Line 2 deliver
$0.8 \cdot (1 - (1 - p(t_1) \cdot p(t_5)) \cdot (1 - p(t_2) \cdot p(t_6)) )$
which can be simplified to $0.3408$.
$\diamond$
\end{example}

\noindent\textbf{Marginal Probabilities via Polynomials.}
For the theoretical analysis of the learning problem presented in Section~\ref{sec:weightLearning},
we next devise an alternative way of computing marginals via polynomial expressions.
%As a preliminary, we start from the observation that the possible worlds over a set of independent base tuples $\mathcal{T}$ must form a probability distribution.
%\begin{proposition} \label{proposition:Distribution} \cite{PDBbook}
%For a probabilistic database ($\mathcal{T}$,$p$) the possible worlds form
%a probability distribution:
%\[
%\sum_{\mathcal{W} \subseteq \mathcal{T}} \underbrace{P(\mathcal{W}, \mathcal{T})}_{\text{via Eq.}~\eqref{eq:ProbPossibleWorld}} = 1
%\]
%\end{proposition}
%\begin{proof}
%We prove the proposition by induction over the cardinality of $\mathcal{T}$.\\
%\underline{Basis i = 1:}
%$\sum_{\mathcal{W} \subseteq \{t\}} P(\mathcal{W}, \{t_1\}) \stackrel{\eqref{eq:ProbPossibleWorld}}{=} p(t) + (1 - p(t)) = 1$ \\
%\underline{Step $(i-1) \rightarrow i$:}
%Let $\mathcal{T} = \{t_1, \dots, t_{i}\}$, where $t_{i}$ is the new tuple.
%\[
%\begin{array}{l}
%	\sum_{\mathcal{W} \subseteq \mathcal{T}} P(\mathcal{W}, \mathcal{T}) \\
%  \stackrel{\eqref{eq:ProbPossibleWorld}}{=} \sum_{\mathcal{W} \subseteq \mathcal{T}} \prod_{t \in \mathcal{W}} p(t) \cdot \prod_{t \in \mathcal{T} \backslash \mathcal{W}} 
%	(1 - p(t)) \\
%  = \underbrace{(p(t_{i}) + (1 - p(t_{i})))}_{=1} 
%	 \cdot
%		\underbrace{%\textstyle{
%			\sum_{\mathcal{W} \subseteq \mathcal{T} \backslash \{t_{i}\}}
%			%}
%			\prod_{t \in \mathcal{W}} p(t) \cdot \prod_{t \in \mathcal{T} \backslash \mathcal{W}}(1 - p(t))}_{=1 \text{ by hypothesis}} \\
%\end{array}
%\]
%\end{proof}
As a preliminary, we reduce the number of terms in Equation~\eqref{eq:lineageConf}'s
sum by considering just tuples $T(\phi)$ that occur in $\phi$
~\cite{PDBbook}.

\begin{proposition} \label{proposition:varsOnly}
We can compute $P(\phi)$ relying on tuples in $T(\phi)$, only,
by writing:
\begin{equation} \label{eq:confPolynomial}
P(\phi) 
 = \sum_{V \subseteq T(\phi), V \models \phi} \underbrace{P(V, T(\phi))}_{\text{via Eq.}~\eqref{eq:ProbPossibleWorld}}
\end{equation}
\end{proposition}
%\begin{proof}
%\[
%\begin{array}{rl}
%P(\phi) & = \sum_{\mathcal{W}\subseteq \mathcal{T}} \mathbf{1}_{\phi}(\mathcal{W}) \cdot P(\mathcal{W}) \\
% & = \left( \sum_{V\subseteq T(\phi)} \mathbf{1}_{\phi}(V) \cdot P(V) \right) \cdot \underbrace{\left( \textstyle{\sum_{V \subseteq (\mathcal{T}\backslash T(\phi))}} P(V) \right)}_{=1 \text{ by Prop.~\ref{proposition:Distribution}}} \\
%\end{array}
%\]
%\end{proof}
Equation~\eqref{eq:confPolynomial} expresses $P(\phi)$ as a polynomial.
Its terms are defined by Equation~\eqref{eq:ProbPossibleWorld}, and the variables
are $p(t)$ for $t \in T(\phi)$. The polynomial's degree is limited as follows.
\begin{corollary} \label{corr:degree}
A lineage formula $\phi$'s marginal probability $P(\phi)$ can be
expressed by a multi-linear polynomial over variables $p(t)$, for $t \in T(\phi)$, with a degree of at most $|T(\phi)|$.
\end{corollary}
\begin{proof}
By inspecting Proposition~\ref{proposition:varsOnly}, we note that the sum ranges
over subsets of $T(\phi)$ only, hence each term has a degree of at most $|T(\phi)|$.
\end{proof}
\begin{example}
Considering the lineage formula
$\phi \equiv t_1 \lor t_2$, the occurring tuples
are $T(\phi) = \{t_1, t_2\}$.
Then, it holds that $\{t_1, t_2\} \models \phi$,
$\{t_1\} \models \phi$, and
$\{t_2\} \models \phi$.
Hence, we can write
$P(\phi) =
p(t_1) \cdot p(t_2) +
p(t_1) \cdot (1- p(t_2)) +
(1 - p(t_1)) \cdot p(t_2)$. Thus, $P(\phi)$ is a polynomial
over the variables $p(t_1)$, $p(t_2)$ and has
degree $2 = |T(\phi)| = |\{t_1,t_2\}|$. $\diamond$
\end{example}

%% file: problem.tex
\section{Learning Problem}
\label{sec:weightLearning}
We now move away from the case where the probability values of all base tuples are
known. Instead, we intend to learn the unknown probability values of (some of)
these tuples (e.g.\ of $t_5$--$t_9$ in Example~\ref{ex:intro}). 
More formally, for a tuple-independent probabilistic database $(\mathcal{T}, p)$, we consider $\mathcal{T}_l \subseteq \mathcal{T}$ to be the set of base tuples for which we learn their probability values. 
That is, initially $p(t)$ is unknown for all
$t \in \mathcal{T}_l$.
Conversely, $p(t)$ is known and fixed for all $t \in \mathcal{T} \backslash \mathcal{T}_l$.
To be able to complete $p(t)$, we are given labels in the form of pairs
$(\phi_i,l_i)$, each containing a
lineage formula $\phi_i$ (i.e., a query answer) and its desired marginal probability $l_i$.
We formally define the resulting learning problem as follows.

\begin{definition} \label{def:learningProblem}
We are given a probabilistic database $(\mathcal{T},p)$, 
a set of tuples $\mathcal{T}_l \subseteq \mathcal{T}$ with unknown probability values $p(t_l)$ %, t_l \in \mathcal{T}_l$,
and a multi-set of given labels $\mathcal{L} = \langle(\phi_1,l_1),\dots,$ $(\phi_n,l_n)\rangle$,
where each $\phi_i$ is a lineage formula over $\mathcal{T}$ and each
$l_i \in [0,1] \subset \mathbb{R}$ is a marginal probability for $\phi_i$.
Then, the {\em learning problem} is defined as follows:
\[
\begin{array}{l}
\text{Determine:}~~p(t_l) \in [0,1] \subset \mathbb{R}~\text{for all}~t_l\in \mathcal{T}_l \\
\text{such that:}~~~P(\phi_i) = l_i~\text{for all}~(\phi_i, l_i) \in \mathcal{L} \\
\end{array}
\]
\end{definition}
Intuitively, we aim to set the probability values of the base tuples
$t_l \in \mathcal{T}_l$ such that the labeled lineage formulas $\phi_i$
yield the marginal probability $l_i$. We want to remark that probability values of tuples in
$\mathcal{T} \backslash \mathcal{T}_l$ remain unaltered.
Also, we note that the Boolean labels {\em true} and {\em false} can be
represented as $l_i = 0.0$ and $l_i = 1.0$, respectively. Hence, Boolean labels resolve to a
special case of Definition~\ref{def:learningProblem}'s labels.

\begin{example} \label{ex:learningProblem} 
Formalizing Example~\ref{ex:intro}'s problem setting, we obtain
$\mathcal{T} := \{t_1, \dots, t_9\}$,
$\mathcal{T}_l := \{t_5, \dots, t_9\}$ with labels $((t_1 \land t_5 \land t_8) \lor (t_2 \land t_6 \land t_8), 0.7)$,
and $((t_3 \land t_7 \land t_9), 0.0)$. $\diamond$
\end{example}

Unfortunately, the above problem definition exhibits hard instances.
First, computing $P(\phi_i)$ may be $\#\mathcal{P}$-hard~\cite{DalviDichotomyConjQuerSharpP}, which would
require many Shannon expansions. But even for cases when all $P(\phi_i)$ can be computed in polynomial time (i.e., when Equation~\eqref{eq:confComp} is applicable),
there are combinatorially hard cases of the above learning problem.

\begin{lemma}\label{lemma:npHard}
For a given instance of Definition~\ref{def:learningProblem}'s learning problem, where all
$P(\phi_i)$ with $(\phi_i, l_i) \in \mathcal{L}$ can be computed in polynomial time,
deciding whether there exists a solution to the learning problem is $\mathcal{NP}$-hard.
\end{lemma}
\begin{proof}
We encode the 3-satisfiability problem (3SAT) for a Boolean formula
$\Psi \equiv \psi_1 \land \dots \land \psi_n$ in CNF into
Definition~\ref{def:learningProblem}'s learning problem.
For each variable $X_i \in \mathit{Var}(\Psi)$,
we create two tuples $t_i$, $t'_i$ whose probability values will be learned.
Hence, $2\cdot |\mathit{Var}(\Psi)| = |\mathcal{T}_l| = |\mathcal{T}|$.
Then, for each $X_i$, we add the label
$( (t_i \land t'_i) \lor (\neg t_i \land \neg t'_i), 1.0)$.
The corresponding polynomial
equation $p(t_i)p(t'_i) + (1-p(t_i))(1-p(t'_i)) = 1.0$ has exactly two
possible solutions for $p(t_i), p(t'_i) \in [0,1]$, namely $p(t_i)=p(t'_i)=1.0$
and $p(t_i)=p(t'_i)=0.0$.
Next, we replace all variables $X_i$ in $\Psi$ by their tuple $t_i$.
Now, for each clause $\psi_i$ of $\Psi$, we introduce one label $(\psi_i, 1.0)$.
Altogether, we have $|\mathcal{L}| = |\mathit{Var}(\Psi)| + n$ labels 
for Definition~\ref{def:learningProblem}'s problem. Each labeled
lineage formula $\phi$ has at most three variables, hence $P(\phi)$ takes
at most 8 steps.
Still, Definition~\ref{def:learningProblem} solves 3SAT, where the learned
values of each pair of $p(t_i)$, $p(t'_i)$ (either $0.0$ or $1.0$) correspond
to $X_i$'s truth value for a satisfying assignment of $\Psi$. From this, it follows that the decision problem formulated in Lemma~\ref{lemma:npHard} is $\mathcal{NP}$-hard. 
\end{proof}

%Moreover, from Lemma~\ref{lemma:npHard} it follows
%that finding a solution to Definition~\ref{def:learningProblem}'s learning problem
%is $\mathcal{FNP}$-hard
%(see~\cite{DBLP:books/daglib/0072413}).

Besides computationally hard instances, there might
also be {\em inconsistent} instances of the learning problem. That is, it may be impossible
to define $p : \mathcal{T}_l \rightarrow [0,1]$ such that
all labels are satisfied.
\begin{example} \label{ex:inconsistent}
If we consider $\mathcal{T}_l := \{t_1, t_2\}$ with the labels
$\mathcal{L} := \langle(t_1, 0.2), (t_2, 0.3), (t_1 \land t_2, 0.9)\rangle$,
then it is impossible to fulfill all three labels at the same time.
\end{example}

From a practical point of view, there remain a number of questions regarding
Definition~\ref{def:learningProblem}.
First, how many labels do we need in comparison to the number of tuples for which
we are learning the probability values (i.e., $|\mathcal{L}|$ vs. $|\mathcal{T}_l|$)?
And second, is there a difference in labeling lineage formulas
that involve many tuples or very few tuples (i.e., $|T(\phi_i)|$)?
These questions will be answered by the following theorem.
It is based on Corollary~\ref{corr:degree}'s computation of marginal probabilities via their polynomial representation. We write the learning problem's
conditions $P(\phi_i) = l_i$ as polynomials over variables $p(t_l)$ of the form $P(\phi_i) - l_i$, where $t_l \in \mathcal{T}_l$ and the probability values
$p(t)$ for all $t \in \mathcal{T} \backslash \mathcal{T}_l$ are fixed
and hence represent constants.

\begin{theorem} \label{th:numberOfSolutions}
If the labeling is consistent, Definition~\ref{def:learningProblem}'s
problem instances can be classified as follows:
\squishlist
  \item[1.] If $|\mathcal{L}| < |\mathcal{T}_l|$, the problem has infinitely
	  many solutions.
	\item[2.] If $|\mathcal{L}| = |\mathcal{T}_l|$ and the polynomials $P(\phi_i) - l_i$ have
		common zeros, then the problem has infinitely many solutions.
	\item[3.] If $|\mathcal{L}| = |\mathcal{T}_l|$ and the polynomials $P(\phi_i) - l_i$
		have no common zeros,	then the problem has at most
		$\prod_i |T(\phi_i) \cap \mathcal{T}_l|$ solutions.
	\item[4.] If $|\mathcal{L}| > |\mathcal{T}_l|$, then the polynomials $P(\phi_i) -l_i$ have common zeros,
		thus reducing this to one of the previous cases.
\squishend
\end{theorem}

\begin{proof}
The first case is a classical under-determined system of equations.
In the second case, without loss of generality, there are two polynomials
$P(\phi_i) - l_i$ and $P(\phi_j) - l_j$ with a common zero, say $p(t_k) = c_k$.
Setting $p(t_k) = c_k$ satisfies both $P(\phi_i) - l_i = 0$ and $P(\phi_j) - l_j = 0$,
hence we have $\mathcal{L}' := \mathcal{L} \backslash \langle(\phi_i,l_i),(\phi_j,l_j)\rangle$
and $\mathcal{T}'_l := \mathcal{T}_l \backslash \{t_k\}$ which yields the theorem's
first case again ($|\mathcal{L}'| < |\mathcal{T}'_l|$).
Regarding the third case,  Bezout's theorem~\cite{Dickenstein}, a central result from algebraic geometry, is applicable: for a system of polynomial
equations, the number of solutions (including their multiplicities) over variables in
$\mathbb{C}$ is equal to the product of the degrees of the polynomials.
In our case, the polynomials are $P(\phi_i) - l_i$ with variables
$p(t_l), t_l \in \mathcal{T}_l$. So, according 
to Corollary~\ref{corr:degree} their degree is at most $|T(\phi_i) \cap \mathcal{T}_l|$.
Since our variables
$p(t_l)$ range only over $[0,1] \subset \mathbb{R}$, and
Corollary~\ref{corr:degree} is an upper bound only, $\prod_i|T(\phi_i) \cap \mathcal{T}_l|$ is an upper bound on the number of solutions.
In the fourth case, the system of equations is over-determined, such that redundancies
like common zeros will reduce the problem to one of the previous cases.
\end{proof}

\begin{example} \label{ex:theorem}
We illustrate the theorem by providing examples for each of the four cases.
\squishlist
  \item[1.] In Example~\ref{ex:learningProblem}'s formalization of Example~\ref{ex:intro},
		we have $|\mathcal{T}_l| = 5$ and
		$|\mathcal{L}| = 2$. So, the problem is under-specified and has infinitely many solutions,
		since assigning $p(t_7)=0.0$ enables $p(t_9)$ to take any value
		in $[0,1] \subset \mathbb{R}$.
	\item[2.] We assume $\mathcal{T}_l = \{t_5, t_6, t_7\}$, and
		$\mathcal{L} = \langle(t_5 \land \neg t_6, 0.0), (t_5 \land \neg t_6 \land t_7, 0.0),
		(t_5 \land t_7, 0.0)\rangle$. This results in the equations
		$p(t_5) \cdot (1 - p(t_6)) = 0.0$, $p(t_5) \cdot (1 -p(t_6)) \cdot p(t_7) = 0.0$,
		and $p(t_5) \cdot p(t_7) = 0.0$, where $p(t_5)$ is a common zero to all
		three polynomials.
		Hence, setting $p(t_5) = 0.0$ allows $p(t_6)$ and $p(t_7)$ to take any
		value in $[0,1] \subset \mathbb{R}$.
	\item[3.] Let us consider $\mathcal{T}_l = \{t_7, t_8\}$.
	\squishlist
		\item[(a)] If $\mathcal{L} = \langle(t_7, 0.4), (t_8, 0.7)\rangle$, then there is exactly one solution
		  as predicted by $|T(t_7)| \cdot |T(t_8)| = 1$. 
		\item[(b)] If $\mathcal{L} = \langle(t_7 \land t_8, 0.1), (t_7 \lor t_8, 0.6)\rangle$, then there are
		 two solutions, namely $p(t_7) = 0.2$, $p(t_8) = 0.5$ and $p(t_7) = 0.5$, $p(t_8) = 0.2$.
		 Here, $\prod_i |T(\phi_i) \cap \mathcal{T}_l| = |T(t_7 \land t_8)| \cdot |T(t_7 \lor t_8)| = 4$
		 is an upper bound.
	\squishend
	\item[4.] We extend this example's second case by the label $(t_5, 0.0)$, thus
	  yielding the same solutions but having $|\mathcal{L}| > |\mathcal{T}_l|$.
	$\diamond$
\squishend
\end{example}
In general, a learning problem instance has many solutions, where
Definition~\ref{def:learningProblem} does not specify a precedence, but
all of them are equivalent.
The number of solutions shrinks by adding labels to $\mathcal{L}$, or by labeling
lineage formulas $\phi_i$ that involve fewer tuples in $\mathcal{T}_l$ (thus resulting in a smaller intersection $|T(\phi_i) \cap \mathcal{T}_l|$).
% "Thus" kommt im Satz davor vor!
Hence, to achieve more uniquely specified probabilities
for all tuples $t_l \in \mathcal{T}_l$, in practice
we should obtain the same number of labels as the number of tuples for which we learn their probability values, i.e., $|\mathcal{L}| = |\mathcal{T}_l|$, and label those lineage formulas with fewer tuples in $\mathcal{T}_l$.

\begin{figure*}[ht]
\begin{center} 
  \subfigure[Example~\ref{ex:theorem}: 3(a): Labels]{ \label{fig:theoremSimple}
    \includegraphics[width=0.33\textwidth]{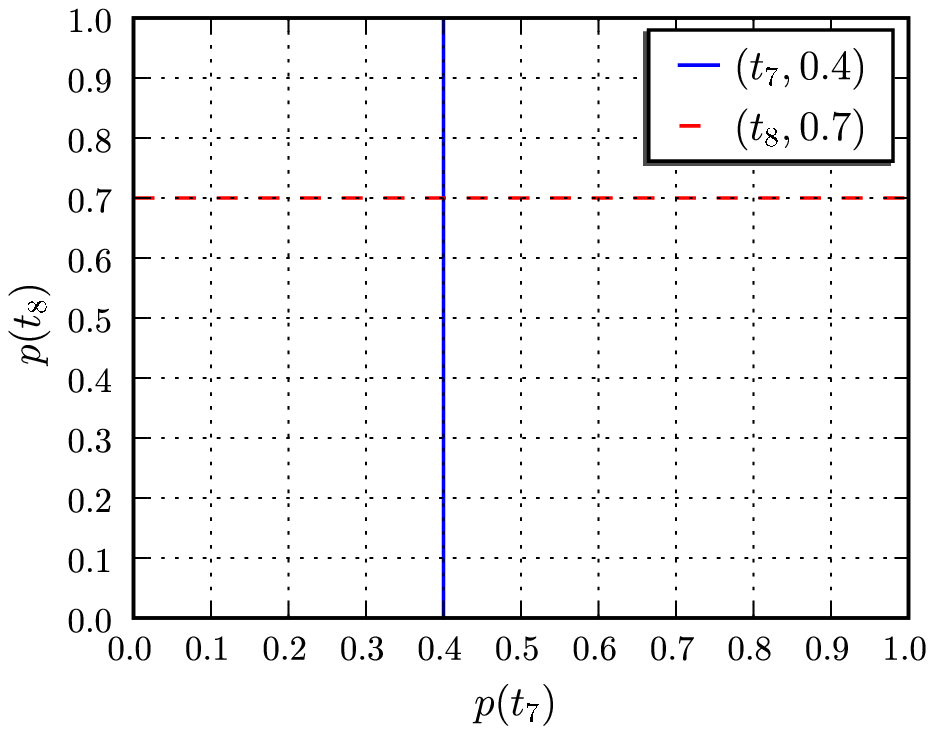}
	}
	\hspace{-6mm}
  \subfigure[Example~\ref{ex:theorem}: 3(b): Labels]{ \label{fig:theoremExample}
    \includegraphics[width=0.33\textwidth]{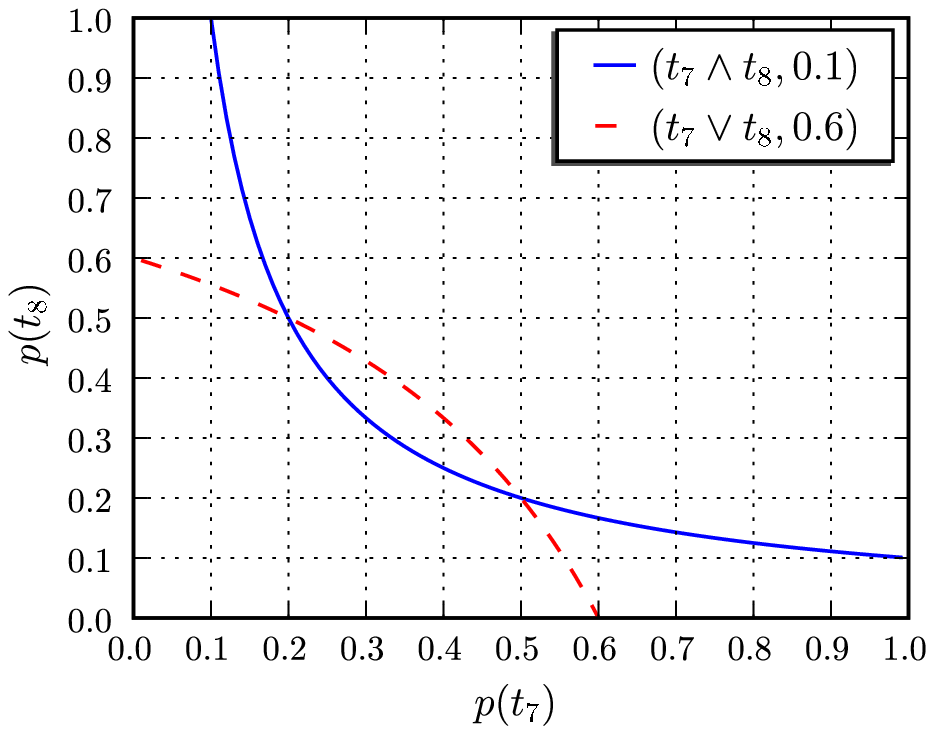}
	}
	\hspace{-6mm}
  \subfigure[Example~\ref{ex:theorem}: 3(b): MSE objective]{ \label{fig:theoremSquared}
    \includegraphics[width=0.33\textwidth]{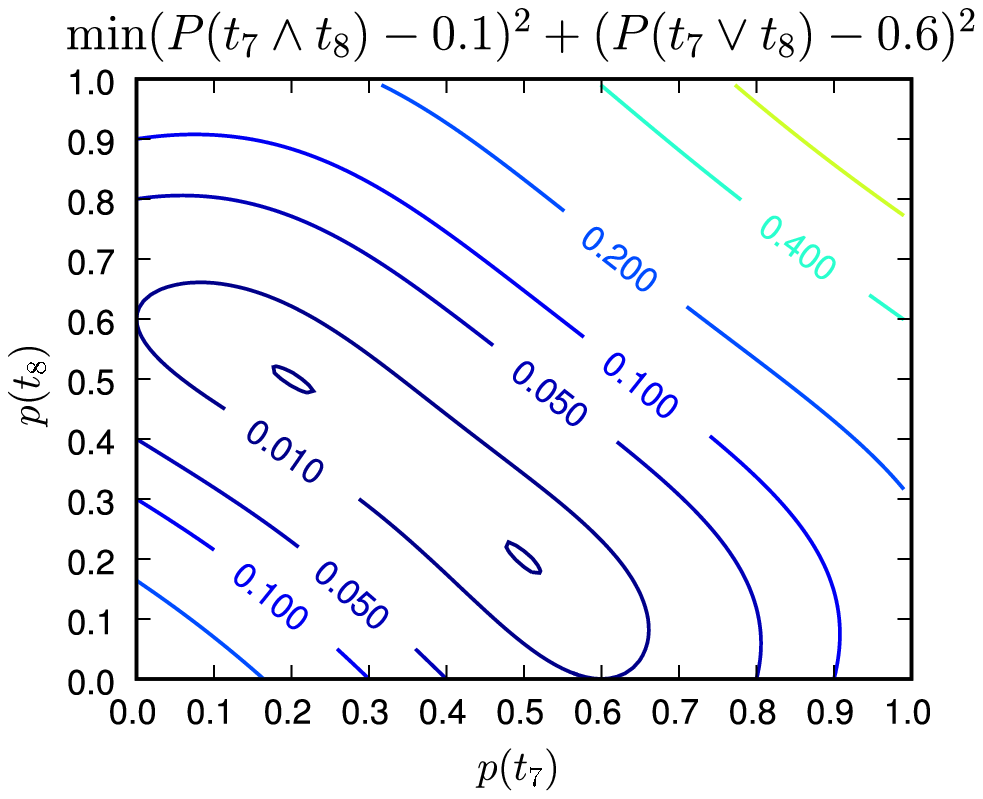}
	}
	\\ \vspace{3mm}
  \subfigure[Example~\ref{ex:logicalObjective}: Logical objective]{ \label{fig:logicalObjective}
    \includegraphics[width=0.33\textwidth]{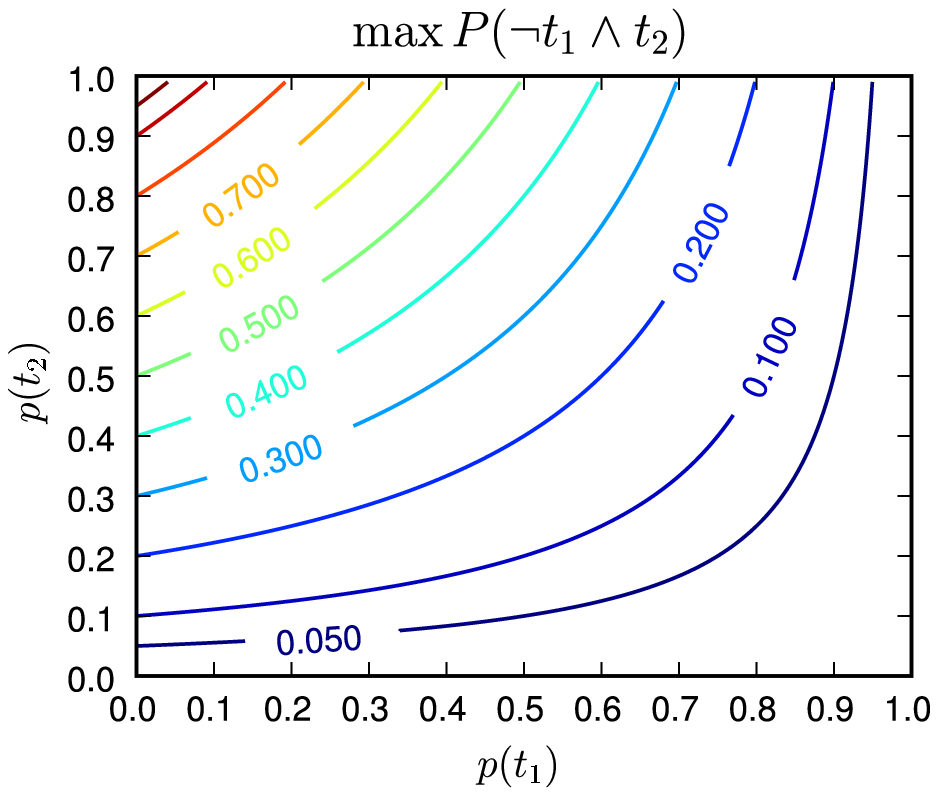}
	}
	\hspace{-6mm}
	\subfigure[Example~\ref{ex:nonStableGradient}: MSE objective]{ \label{fig:squaredOnce}
    \includegraphics[width=0.33\textwidth]{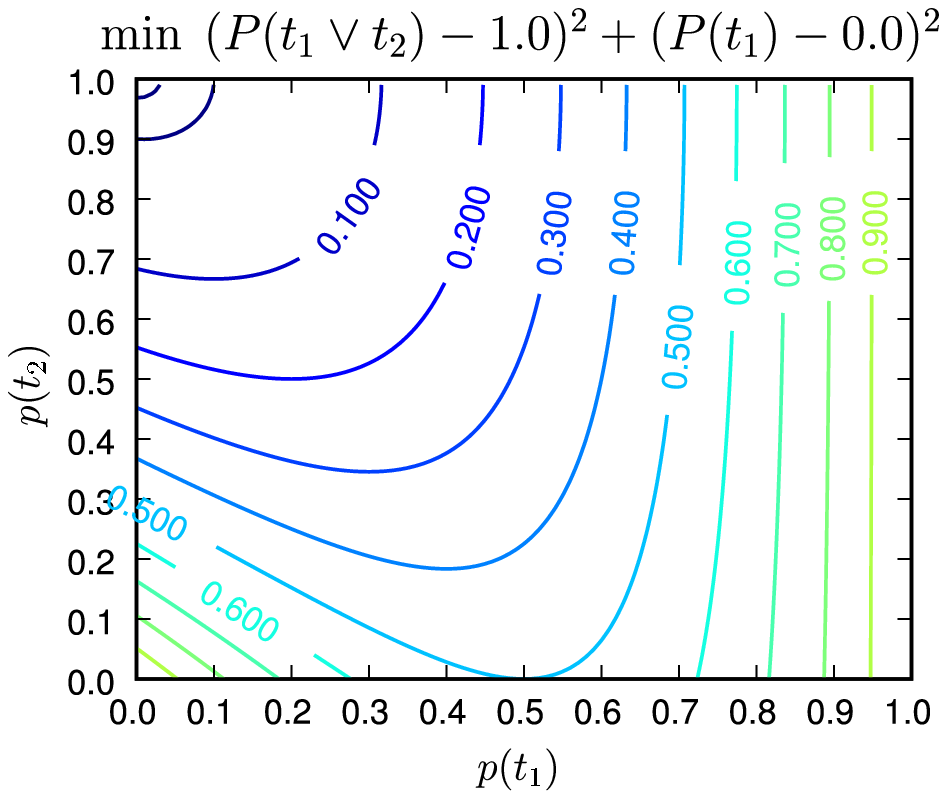}
	}
	\hspace{-6mm}
	\subfigure[Example~\ref{ex:nonStableGradient}: MSE objective, unstable]{ \label{fig:squaredTwice}
    \includegraphics[width=0.33\textwidth]{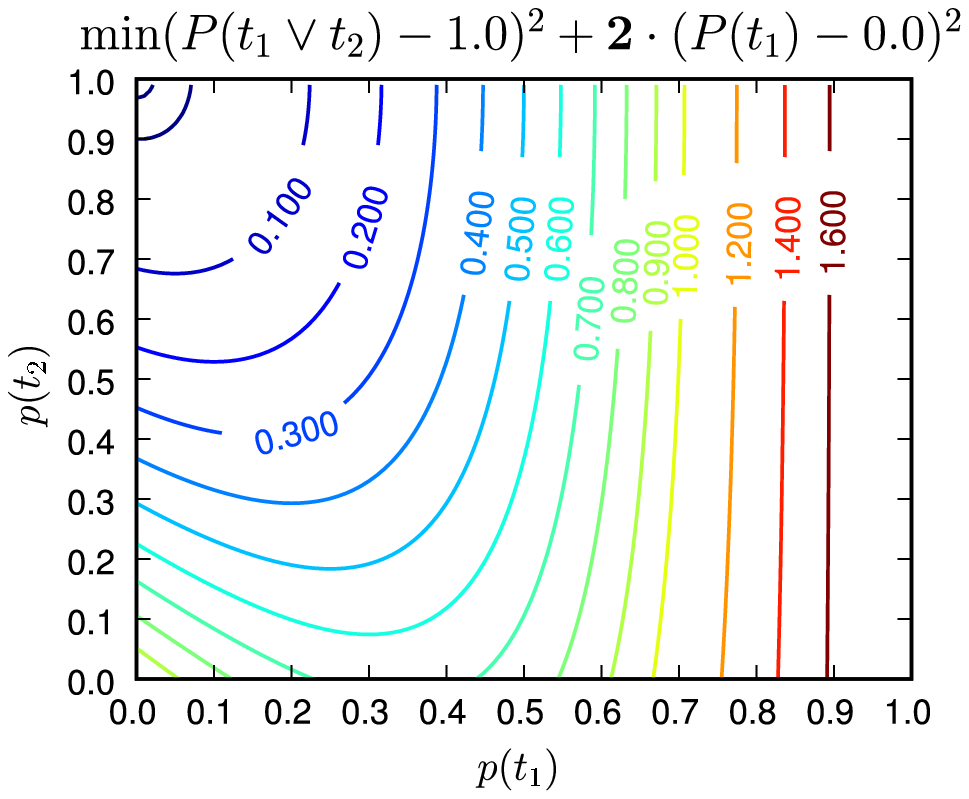}
	}
\end{center}
\vspace{-7mm}
\caption{Visualization of the Learning Problem}
\end{figure*}

Based on algebraic geometry, the learning problem allows for an interesting {\em visual interpretation}.
All possible definitions of probability values for tuples in $\mathcal{T}_l$,
that is $p:\mathcal{T}_l \rightarrow [0,1]$, span the hypercube
$[0,1]^{|\mathcal{T}_l|}$. In Example~\ref{ex:theorem}, Cases 3(a) and 3(b),
the hypercube has two dimensions, namely $p(t_7)$ and $p(t_8)$, as depicted
in Figures~\ref{fig:theoremSimple} and~\ref{fig:theoremExample}.
Hence, one definition of $p$ specifies exactly one point in the hypercube.
Moreover, all definitions of $p$ that satisfy a given label define a
curve (or plane) through the hypercube (e.g., the two labels in Figure~\ref{fig:theoremSimple} define
two straight lines).
Also, the points, in which all labels' curves intersect, represent solutions
to the learning problem (e.g., the solutions of Example~\ref{ex:theorem}, Case 3(b), are the intersections in Figure~\ref{fig:theoremExample}).
If the learning problem is inconsistent, there is no point in which all labels' curves
intersect. Furthermore, if the learning problem has infinitely many solutions,
the labels' curves intersect in curves or planes, rather than points.

%% file: solving.tex
\section{Solving the Learning Problem}
\label{sec:solving}
In the previous section, we formally characterized the learning problem and devised the basic properties of its solutions.
From a visual perspective, Definition~\ref{def:learningProblem} established curves and planes whose intersections represent the solutions (see, e.g., Figure~\ref{fig:theoremExample}).
In this section, we introduce different objective functions that describe surfaces whose optima correspond to these solutions. For instance,
Figure~\ref{fig:theoremExample}'s problem has Figure~\ref{fig:theoremSquared}'s  
surface if we the use mean squared error (MSE) as the objective, which will be defined in this section. Calculating a gradient on such a surface thus allows the application
of an optimization method to solve the learning problem.

\noindent \textbf{Alternative Approaches.}
In general,
based on the polynomial equations, an exact solution to an instance of the learning problem
can be computed in exponential time~\cite{Dickenstein}, which is not
acceptable in a database setting. Also, besides gradient-based
optimization methods, other approaches, such as
expectation maximization~\cite{GuptaEM}, are
possible and represent valuable targets for future work.

\noindent \textbf{Derivative.}
In order to establish a gradient on Definition~\ref{def:learningProblem}'s
conditions, i.e., $P(\phi_i) = l_i$, we introduce the partial derivative
of a lineage formula's
marginal probability $P(\phi)$ with respect to a given tuple $t \in T(\phi)$.
\begin{definition} \label{def:confDerivative}
\cite{SensitivityPDB}
Given a lineage formula $\phi$ and a tuple $t \in T(\phi)$, the partial derivative
of $P(\phi)$ with respect to $p(t)$ is defined as:
\[
\frac{\partial P(\phi)}{\partial p(t)} := P(\phi_{[t\rightarrow\mathit{true}]})-P(\phi_{[t\rightarrow\mathit{false}]})
\]
\end{definition}
Here, $\phi_{[t \rightarrow \mathit{true}]}$ means that all occurrences of $t$ in $\phi$
are replaced by {\em true} (and analogously for {\em false}).
%A formal justification for the above definition is provided in~\cite{SensitivityPDB}.

\begin{example}
Considering the marginal probability $P(t_1 \lor t_2)$ with $p(t_1) = 0.6$,
we determine the partial derivative with respect to $p(t_2)$, that is
$\frac{\partial P(t_1 \lor t_2)}{\partial p(t_2)} = P(t_1 \lor \mathit{true}) - P(t_1 \lor \mathit{false})
= 1.0 - 0.6 = 0.4$.
$\diamond$
\end{example}

\noindent\textbf{Desired Properties.}
Before we define objective functions for solving the learning
problem, we establish a list of desired properties of these (which
we do not claim to be complete). Later, we judge different objectives
based on these properties.
\begin{definition} \label{def:desired}
An objective function to the learning problem should
satisfy the following three {\em desired properties}:
\squishlist
  \item[1.] All instances of Definition~\ref{def:learningProblem}'s learning problem
	      can be expressed, including inconsistent ones.
	\item[2.] If all $P(\phi_i)$ are computable in polynomial time, then also
	      the objective is computable in polynomial time.
	\item[3.] The objective is stable, that is
	      $\mathcal{L} := \langle(\phi_1, l_1), \dots,$ $(\phi_n,l_n)\rangle$
        and $\mathcal{L} \cup \langle(\phi_i', l_i)\rangle$ with
				$\phi_i' \equiv \phi_i$, $(\phi_i, l_i) \in \mathcal{L}$
				define the same surface.
\squishend
\end{definition}
Here, the first case ensures that the objective can be applied to all instances
of the learning problem. We insist on inconsistent instances, because they
occur often in practice (see Figure~\ref{table:datasets}).
The second property restricts a blow-up in computation,
which yields the following useful characteristic: if we can compute $P(\phi)$
for all labels, e.g., for labeled query answers, then we can also compute
the objective function. Finally, the last of the desiderata reflects an objective
function's ability to detect dependencies between labels. Since
$\phi_i \equiv \phi_i'$ both $\mathcal{L}$ and $\mathcal{L} \cup \langle (\phi_i',l_i) \rangle$
allow exactly the same solutions, the surface should be the same.
Unfortunately, including convexity of an objective as an
additional desired property is not possible. For example Figure~\ref{fig:theoremExample}
has two disconnected solutions, which induce at least two optima,
thus prohibiting convexity.
In the following, we establish two objective functions, which behave
very differently with respect to the desired properties.

\noindent\textbf{Logical Objective.}
If we restrict the learning problem's probability labels to $l_i \in \{0.0, 1.0\}$,
we can define a objective function based on computing marginals as follows.
\begin{definition} \label{def:logicalObjective}
Let an instance of Definition~\ref{def:learningProblem}'s learning problem
be given by a probabilistic database $(\mathcal{T},p)$, tuples
with unknown probability values $\mathcal{T}_l \subseteq \mathcal{T}$, and labels $\mathcal{L} = \langle (\phi_1, l_1),\dots,$ $(\phi_n, l_n)\rangle$
such that all $l_i \in \{0.0, 1.0\}$.
Then, the {\em logical objective} is formulated as:
\begin{equation} \label{eq:booleanOnly}
\mathit{Logical}(\mathcal{L},p) := P \left(
\bigwedge_{(\phi_i,l_i) \in \mathcal{L}, l_i = 1.0} \!\!\!\!\!\!\phi_i
~
\land
\bigwedge_{(\phi_i,l_i) \in \mathcal{L}, l_i = 0.0} \!\!\!\!\!\!\!\!\!\!\neg \phi_i \right)
\end{equation}
\end{definition}
The above definition is a maximization problem, and its global optima are identified by
$\mathit{Logical}(\mathcal{L},p) = 1.0$.
Moreover, from Definition~\ref{def:confDerivative}, we may obtain its derivative.
\begin{example} \label{ex:logicalObjective}
Let $\mathcal{T} = \mathcal{T}_l := \{t_1, t_2\}$
and $\mathcal{L} := \langle(t_1 \lor t_2, 1.0), (t_1, 0.0)\rangle$ be given.
Then, $\mathit{Logical}(\mathcal{L},p)$ is instantiated as
$P((t_1 \lor t_2) \land \neg t_1) = P(\neg t_1 \land t_2)$.
Visually, this defines a surface
whose optimum lies in $p(t_1) = 0.0$ and $p(t_2) = 1.0$, as shown in
Figure~\ref{fig:logicalObjective}.
$\diamond$
\end{example}
With respect to Definition~\ref{def:desired}, the third desired property
is fulfilled, as $P(\phi_i' \land \phi_i) = P(\phi_i)$. Hence, the logical
objective's surface, shown for instance in Figure~\ref{fig:logicalObjective},
is never altered by adding equivalent labels.
Still, the first property is not given, since the probability labels
are restricted to $l_i \in \{0.0, 1.0\}$ and inconsistent problem
instances collapse Equation~\eqref{eq:booleanOnly} to $P(\mathit{false})$, thus rendering
the objective non-applicable. Also, the second property is violated, because
in the spirit of Lemma~\ref{lemma:npHard}'s proof, we can construct an instance
where each label's $P(\phi_i)$ on its own is computable in polynomial time, whereas
the computation of the marginal probability for Equation~\eqref{eq:booleanOnly} is $\mathcal{\#P}$-hard.

\noindent\textbf{Mean Squared Error Objective.}
Another approach, which is common in machine learning, lies in using the mean squared error (MSE)
to define the objective function.
\begin{definition} \label{def:MSE}
Let an instance of Definition~\ref{def:learningProblem}'s
learning problem be given by a probabilistic database $(\mathcal{T}, p)$, 
tuples with unknown probability values $\mathcal{T}_l \subseteq \mathcal{T}$,
and labels $\mathcal{L} = \langle(\phi_1, l_1), \dots,$ $(\phi_n, l_n)\rangle$.
Then, the {\em mean squared error objective function} is formulated as:
\[
\mathit{MSE}(\mathcal{L},p) := \frac{1}{|\mathcal{L}|}\sum_{(\phi_i, l_i) \in \mathcal{L}} (P(\phi_i) - l_i)^2
\]
Moreover, its partial derivative with respect to the tuple's probability value $p(t)$
is:
\[
\frac{\partial \mathit{MSE}(\mathcal{L},p)}{\partial p(t)} :=
\frac{1}{|\mathcal{L}|} \sum_{(\phi_i, l_i) \in \mathcal{L},
                              t \in T(\phi_i)}
\!\!\!\!\!2 \cdot (P(\phi_i) - l_i) \cdot \underbrace{\frac{\partial P(\phi_i)}{\partial p(t)}}_{\text{Def.}~\ref{def:confDerivative}}
\]
\end{definition}
The above formulation is a minimization problem whose solutions have $0.0$ as
the objective's value.
\begin{example}
Example~\ref{ex:theorem}, Case 3(b), is visualized in
Figure~\ref{fig:theoremExample}. The corresponding
surface induced by the MSE objective is
depicted in Figure~\ref{fig:theoremSquared} and
has its minima at the learning problem's solutions. $\diamond$
\end{example}
Judging the above objective by means of Definition~\ref{def:desired},
we realize that the first property is met, as there are no restrictions
on the learning problem, and inconsistent instances can be tackled (but deliver
objective values larger than zero). Furthermore, since the $P(\phi_i)$'s occur
in separate terms of the objective's sum, the second desired property is maintained.
However, the third desired property is violated, as illustrated by the following example.
\begin{example} \label{ex:nonStableGradient}
In accordance to Example~\ref{ex:logicalObjective} and Figure~\ref{fig:logicalObjective}, we set
$\mathcal{T} = \mathcal{T}_l := \{t_1, t_2\}$ and
$\mathcal{L} := \langle(t_1 \lor t_2, 1.0),$ $(t_1, 0.0)\rangle$.
Then, the MSE objective defines the surface in 
Figure~\ref{fig:squaredOnce}. However, if we replicate
the label $(t_1, 0.0)$, thus resulting
in Figure~\ref{fig:squaredTwice} (note the ``times two'' in
the objective), its surface becomes steeper along the $p(t_1)$-axis,
but has the same minimum.
Thus, MSE's surface is not stable. Instead, it becomes more
ill-conditioned~\cite{OptimizationBook}.
$\diamond$
\end{example}

\noindent \textbf{Discussion.} Both the logical objective and the
MSE objective have optima exactly at the solutions of
Definition~\ref{def:learningProblem}'s learning problem.
With respect to Definition~\ref{def:desired}'s desired properties, we summarize
the behavior of both objectives in the following table:
\vspace{-3mm}
\[
\begin{array}{r|c|c|c}
 & \multicolumn{3}{c}{\text{Properties}} \\
\text{Objective} & 1. & 2. & 3. \\
 \hline
\text{Logical} & \times & \times & \checkmark \\
\text{MSE} & \checkmark & \checkmark & \times \\
\end{array}
\]
The two objectives satisfy opposing desired properties, and it is certainly
possible to define other objectives behaving similarly to one of them.
Unfortunately, there is little hope for an objective that is adhering to all three properties.
The second property inhibits computational hardness. However,
Lemma~\ref{lemma:npHard} and
the third property's logical tautology checking (i.e.,
$\models \phi_i \leftrightarrow \phi_i'$, which is co-$\mathcal{NP}$-complete)
require these. In this regard the logical objective addresses both computationally
hard problems by computing marginals, whereas the MSE objective avoids them.

In the remainder of the paper, we will favor the
MSE objective, as it is more practical.
In reality, many learning problem instances are inconsistent or have
non-Boolean labels (see Figure~\ref{table:datasets}),
and Equation~\eqref{eq:booleanOnly}'s marginal computations
are often too expensive (see Section~\ref{sec:exp:objectives}).

%% file: algorithms.tex
\section{Learning Algorithm}
\label{sec:algorithm}
Given a learning problem's 
surface (see, e.g., Figure~\ref{fig:theoremSquared}), as it is defined by 
the choice of the objective function, this section's learning algorithm determines how to
move over this surface in order to reach an optimum, that is, to find a solution to the learning problem.

\noindent \textbf{Learning Algorithm.}
Our learning algorithm is based on stochastic gradient
descend (SGD)~\cite{NIPS2007_726}, which we demonstrate to scale to instance sizes with
millions of tuples and hundreds of thousands of labels (see Section~\ref{sec:exp:scalability}).
It is initialized at a random point and
repeatedly moves into the direction of a partial derivative until convergence.
Visually, we start at a random point (e.g., somewhere in Figure~\ref{fig:theoremSquared}), and then in each step we move in 
parallel to an axis (e.g., $p(t_1)$ or $p(t_2)$), until we reach an optimum.

In Algorithm~\ref{alg:sgd} $\mathit{best}$, represents the
objective's best known value,
where $p$ holds the corresponding probability values of tuples in $\mathcal{T}_l$.
Also, $\eta_l$ is the learning rate, which exists and may differ for each tuple
in $\mathcal{T}_l$.
Line~\ref{line:while}'s loop is executed until convergence to the absolute error bound of
$\epsilon_\mathit{abs}$.
Then, Line~\ref{line:shuffle} shuffles the order of $\mathcal{T}_l$'s tuples for 
the inner loop of Line~\ref{line:whileSequence}. Within each iteration,
Line~\ref{line:move} updates one tuple's probability value, which yields the
updated definition $p'$ of $p$. If $p'$ is an improvement
over $p$ with respect to
the objective (as verified in Line~\ref{line:improvement?}),
we assign $p'$ to $p$ and double the tuple's learning rate $\eta_l$.
Otherwise, $p'$ is discarded, and the learning rate $\eta_l$ is halved.
%Increasing and decreasing the learning rate depending on the improvement
%is sometimes refered to as bold driver heuristic.
\begin{algorithm}
\small
\caption{Learning($(\mathcal{T},p),\mathcal{T}_l,\mathcal{L},\epsilon_\mathit{abs}$)} \small \label{alg:sgd}
\begin{algorithmic}[1]
  \Require Probabilistic database $(\mathcal{T},p)$, tuples $\mathcal{T}_l$ to
	  learn the probability values for,
	  labeled lineage formulas $\mathcal{L}$, error bound $\epsilon_\mathit{abs}$
  \Ensure $p$ with learned probability values, $\mathit{best}$ value of objective
	\State $\forall t_l \in \mathcal{T}_l: p(t_l) := \mathit{Rand}(0,1)$ \Comment{Random initialization} \label{line:randomStart} \label{line:initRandom}
	\State $\forall t_l \in \mathcal{T}_l: \eta_l := 1.0$ \Comment{Per-tuple learning rate}
	\State $\mathit{best} := \mathit{MSE}(\mathcal{L}, p)$ \Comment{Definition~\ref{def:MSE}} \label{line:initBest}
	\While{$\mathit{best} > \epsilon_\mathit{abs}$} \label{line:while}
		\State $\mathit{sequence} := \mathit{Shuffle}(\mathcal{T}_l)$ \Comment{Permuted sequence} \label{line:shuffle}
		\While{$\neg \mathit{IsEmpty}(\mathit{sequence})$} \label{line:whileSequence}
			\State $t_l := \mathit{Pop}(\mathit{sequence})$ \Comment{Get first element}
			\State $p'(t_l) := p(t_l) - \eta_l \cdot \frac{\partial \mathit{MSE}(\mathcal{L},p)}{\partial p(t_l)}$ \Comment{Definition~\ref{def:MSE}} \label{line:move}
			\State $p' := p \cup \{p'(t_l)\}$
			\State $\mathit{newVal} := \mathit{MSE}(\mathcal{L},p')$ \Comment{Definition~\ref{def:MSE}} \label{line:newVal}
			\If{$\mathit{newVal} < \mathit{best}$} \label{line:improvement?}
				\State $\eta_l := 2 \cdot \eta_l$ \Comment{Increase $t_l$'s learning rate}
				\State $p := p'$ \Comment{Keep new value of $p(t_l)$}
				\State $\mathit{best} := \mathit{newVal}$
			\Else
				\State $\eta_l := \frac{1}{2} \cdot \eta_l$  \Comment{Decrease $t_l$'s learning rate}
			\EndIf
		\EndWhile
	\EndWhile
  \State \Return $p$, $\mathit{best}$
\end{algorithmic}
\end{algorithm} \normalsize

\begin{example}
We execute Algorithm~\ref{alg:sgd} on Figure~\ref{fig:squaredOnce}'s example.
Following Definition~\ref{def:MSE} the corresponding partial derivatives are:
\begin{equation} \label{eq:squaredOnceDerivatives}
\begin{array}{rl}
\frac{\partial \mathit{MSE}}{\partial p(t_1)} := 
&\!\!\!\! (P(t_1 \lor t_2) - 1.0) \cdot (P(\mathit{true} \lor t_2) - P(\mathit{false} \lor t_2)) \\
&\!\!\!\! + (P(t_1) - 0.0) \cdot (P(\mathit{true}) - P(\mathit{false})) \\
\frac{\partial \mathit{MSE}}{\partial p(t_2)} := 
&\!\!\!\! (P(t_1 \lor t_2) - 1.0) \cdot (P(t_1 \lor \mathit{true}) - P(t_1 \lor \mathit{false})) \\
\end{array}
\end{equation}
Assuming that Line~\ref{line:initRandom} delivers $p(t_1) = 0.7$ and $p(t_2) = 0.5$,
we get $\mathrm{best} = (-0.15)^2 + (0.7)^2 \approx 0.512$ in Line~\ref{line:initBest}.
If $\epsilon_\mathit{abs} = 0.01$ we enter Line~\ref{line:while}'s loop, where Line~\ref{line:shuffle}
randomly orders $t_2$ before $t_1$. Then, $p(t_2)$'s partial derivative evaluates as follows
$\frac{\partial \mathit{MSE}}{\partial p(t_2)}\big|_{(0.7, 0.5)} = (0.85 - 1.0) \cdot (1.0 - 0.7)
= -0.055$.
Since $\eta_2 = 1.0$, we get $p'(t_2) = 0.5 - (-0.055) = 0.555$ in Line~\ref{line:move}.
Hence, in Line~\ref{line:newVal}, $\mathrm{newVal} = (-0.1335)^2 + 0.7^2 \approx 0.508$.
As $0.508 < 0.512$, Line~\ref{line:improvement?}'s condition turns true,
such that we get $\eta_2 = 2.0$, $p(t_1) = 0.7$, $p(t_2) = 0.555$
and $\mathrm{best} = 0.508$. Hence, in further iterations
the increased $\eta_2$ speeds up movements along $p(t_2)$'s partial derivative.
$\diamond$
\end{example}

\noindent \textbf{Tackling MSE's Instability.}
The disadvantage of the MSE objective is that it does not satisfy
Definition~\ref{def:desired}'s third desired property. We argue,
that Algorithm~\ref{alg:sgd} counters to some extent the instability, which
we illustrate by the following example.
\begin{example} \label{ex:gradientChange}
Let us evaluate the gradient of Figures~\ref{fig:squaredOnce}
and \ref{fig:squaredTwice} in the point $p(t_1) = p(t_2) = 0.5$.
Following Equation~\eqref{eq:squaredOnceDerivatives}, we obtain
%$\frac{\partial \mathit{MSE}}{\partial p(t_1)}\big|_{(0.5,0.5)} = -0.25 \cdot 0.5 + 0.5 \cdot 1.0 = 0.375$
%and
%$\frac{\partial \mathit{MSE}}{\partial p(t_2)}\big|_{(0.5,0.5)} = -0.25 \cdot 0.5 = -0.125$
the gradient $(0.375, -0.125)$
for Figure~\ref{fig:squaredOnce}.
Analogously, Figure~\ref{fig:squaredTwice} has $(0.875, -0.125)$.
Even although both figures show the same minimum, the gradients
differ heavily in $p(t_1)$'s partial derivative.
$\diamond$
\end{example}
Inspecting the above example, we note that the gradient
is indeed affected, but each partial derivative on its own
points into the correct direction, i.e.\ increasing $p(t_2)$ and
decreasing $p(t_1)$. Hence, weighting the partial derivatives
can counter the effect. We achieve this by keeping one learning rate $\eta_l$
per tuple and adapting them during runtime. In Section~\ref{sec:exp:gradientBased}, we
empirically show a superior convergence over a global learning rate.
Previously, the authors of~\cite{WeightLearningMLN}
also reported speed ups in ill-conditioned
instances by introducing separate learning rates per dimension.

\noindent \textbf{Implementation Issues.}
In this paragraph, we briefly describe four implementation issues, which were
omitted in Algorithm~\ref{alg:sgd} for presentation purposes.
First, Line~\ref{line:while}'s absolute error bound
is inconvenient, because the optima of inconsistent learning problem
instances have an MSE value larger than $0.0$. Therefore, we
use both an absolute error bound $\epsilon_\mathit{abs}$ and 
a relative error bound $\epsilon_\mathit{rel}$. Second,
since their marginal probabilities $P(\phi_i)$ are repeatedly computed,
it is beneficial to preprocess the lineage formulas $\phi_i$, e.g.\ 
by compiling them to OBDDs~\cite{JhaCompiling},
or by flattening them via a few targeted Shannon expansions~\cite{TPDBdylla}, the latter of which we also apply in this work.
Next, Line~\ref{line:move} might yield a probability value that exceeds
the interval $[0,1]$, which we counter by the {\em logit} function.
It defines a mapping from probability values in $[0,1]$ to
$\mathbb{R} \cup \{\pm\infty\}$.
\begin{definition}\label{def:logit}
The logit function transforms a probability $p \in [0,1]$
to a weight $w \in \mathbb{R} \cup \{\pm\infty\}$ as follows:
\[
w = \ln \frac{p}{1.0 - p} \qquad p = \frac{1}{1 + \exp(-w)}
\]
\end{definition}
\begin{example}
If $p = 0.5$, then $w = 0.0$. Also $p = 1.0$ implies $w = +\infty$,
whereas $p = 0.0$ yields $w = -\infty$. $\diamond$
\end{example}
Hence, we calculate with weights in $\mathbb{R} \cup \{\pm\infty\}$, rather than on
probability values in $[0,1]$.
Finally, if two tuples $t_l, t_l' \in \mathcal{T}_l$ are disjoint with respect
to the labels' lineage formulas they occur in, that is
$\{\phi_i~|~(\phi_i, l_i) \in \mathcal{L}, t_l \in T(\phi_i), t_l'\in T(\phi_i)\} = 
\emptyset$, then their probability values can be updated in parallel.

\noindent \textbf{Alternative Approaches.}
Due to its various applications, there is an entire zoo of
gradient-based optimization techniques~\cite{OptimizationBook}.
Approaches, such as Newton's method, which are based on the Hessian,
do not to scale to database-like instance
sizes. This disadvantage is circumvented by Quasi-Newton methods, for
instance limited-memory Broyden-Fle\-tcher-Goldfarb-Shanno (L-BFGS)~\cite{OptimizationBook},
which estimates the Hessian. We empirically compare our approach to L-BFGS and 
plain gradient descent in Section~\ref{sec:exp:gradientBased}.

\noindent \textbf{Algorithm Properties.}
Algorithm~\ref{alg:sgd} comes with three properties, which
we share with alternative approaches we are aware of, including 
other gradient-based methods and expectation maximization~\cite{GuptaEM}.
First, the algorithm is {\em non-deterministic}, which is 
caused by Lines~\ref{line:randomStart} and~\ref{line:shuffle}.
Second, gradient-based optimization methods, including Algorithm~\ref{alg:sgd},
can get stuck in {\em local optima}, which is nevertheless
hard to avoid in non-convex problems. In this regard, the non-determinism is a potential advantage,
since restarting the algorithm will yield varying solutions, thus increasing
the chance for finding a global optimum.
Finally, the solutions returned by Algorithm~\ref{alg:sgd} for the MSE objective
are {\em not exact}, but rather very close to an optimum, which however can
be controlled by the error bounds $\epsilon_\mathit{rel}$ and $\epsilon_\mathit{abs}$.
Due to space constraints, experiments on these aspects are available in the supplementary
material\textsuperscript{\ref{footnote:supplementary}}.

%% file: relatedProblems.tex
\section{Extensions \& Applications}
\label{sec:relatedProblems}
In this section, we briefly investigate how the learning problem
can be extended by priors (Section~\ref{sec:priors}), how it relates to conditioning PDBs via constraints
(Section~\ref{sec:constraintsAndLearning}), how these constraints can be employed to update or clean PDBs (Section~\ref{sec:updating}), and how it relates to handling incompleteness in databases
(Section~\ref{sec:incompleteAndLearning}).

\subsection{Priors}
\label{sec:priors}
%From a Bayesian perspective, we can refer to $p(t_l), t_l \in \mathcal{T}_l$ before
%learning as prior, whereas the solution to the learning
%problem $p(t_l)$ can be viewed as posterior. Hence, in the previous
%sections the prior was undefined.
%Acquiring labeled lineage formulas from humans can be expensive,
%which we can sidestep by using priors.

In order to explicitly incorporate preferences in the form of prior probabilities of base tuples $t_l \in \mathcal{T}_l$ into our learning objective (instead of just considering them to be ``unknown''), we can extend Definition~\ref{def:MSE}'s MSE objective
%by informative priors for $\mathcal{T}_l$'s probability values 
as follows.
\begin{definition} \label{def:prior}
Given a function $\mathit{prior}: \mathcal{T}_l \rightarrow [0,1] \subset \mathbb{R}$,
Definition~\ref{def:MSE}'s MSE objective function can
be extended to
\[
\frac{c}{|\mathcal{L}|} \cdot \!\! \sum_{(\phi_i, l_i) \in \mathcal{L}} (P(\phi_i) - l_i )^2
+ \frac{1-c}{|\mathcal{T}_l|} \cdot \!\! \sum_{t_l \in \mathcal{T}_l} (P(t_l) - \mathit{prior}(t_l))^2
\]
where $c \in [0,1]$ is a constant. 
\end{definition}
Utilizing $c$, we can control the trade-off between the impact of the lineage labels and the $prior$ function.

\noindent \textbf{Expressiveness.}
Definition~\ref{def:prior} is not more general
than the original MSE objective.
We can express priors in Definition~\ref{def:MSE} by creating a label $(t_l, \mathit{prior}(t_l))$  for each tuple $t_l \in \mathcal{T}_l$, which then produces $\sum_{t_l \in \mathcal{T}_l} (P(t_l) - \mathit{prior}(t_l))^2$ also in Definition~\ref{def:MSE}'s objective. The coefficients preceding the sums can be emulated by replicating labels in $\mathcal{L}$. Thus, priors are a special case of lineage labels. %From a Bayesian perspective, 
% the difference between a prior and a posterior probability of a tuple becomes evident once we {\em condition}~\cite{KochConditioningPDB} the probabilistic database onto constraints (see Subsection~\ref{sec:constraintsAndLearning}). %In absence of constraints the 
%we can refer to $p(t_l), t_l \in \mathcal{T}_l$ before
%learning as prior, whereas the solution to the learning
%problem $p(t_l)$ can be viewed as posterior. Hence, in the previous
%sections the prior was undefined.

\subsection{Constraints}
\label{sec:constraintsAndLearning}
%The consistency of databases is often enforced via constraints,
%where {\em conditioning}~\cite{KochConditioningPDB} explored the problem
%in the PDB setting. We investigate the relationship between both
%problems.

\noindent\textbf{Conditioning by Learning.}
Considering constraints in the form of propositional formulas over
a probabilistic data\-base's tuples, we can
encode each constraint $\phi_i$ with the label $(\phi_i, 1.0)$
into an instance of the learning problem.
%\begin{example}
%In Example~\ref{ex:intro} we can require the $\mathit{BornIn}$ relation
%to be functional, since people are born in only one place.
%Hence, $\mathit{BornIn}(\mathit{Spielberg}, \mathit{Cinncinati})$ and
%$\mathit{BornIn}(\mathit{Spielberg},$ $\mathit{LosAngeles})$ can not
%co-exist, which we express via their lineage formulas as
%$\neg (t_3 \land t_7 \land t_8) \lor \neg (t_4 \land t_7 \land t_9)$.
%If we label this formula by $1.0$, we can enforce it during learning. $\diamond$
%\end{example}

\begin{lemma}
Given a probabilistic database $(\mathcal{T},p)$ and constraints in the
form of propositional formulas $\phi_1, \dots, \phi_n$ over $\mathcal{T}$, whose
conjunction $\phi_1 \land \dots \land \phi_n$ is satisfiable.
Then, if we create a learning problem instance by setting
$\mathcal{T}_l := \mathcal{T}$ and $\mathcal{L} := \langle(\phi_1, 1.0), \dots, (\phi_n, 1.0)\rangle$,
its solution $p'$ conditions the probabilistic database $(\mathcal{T},p)$ with
respect to $\phi_1, \dots, \phi_n$. Hence, for a propositional query $\psi$ over $(\mathcal{T},p')$ it
holds, that:
\[
P(\psi~|~\phi_1 \land \dots \land \phi_n) = P(\psi)
\]
\end{lemma}
\begin{proof}
We observe that in the learning problem's solution $p'$, we get
$P(\phi_1 \land \dots \land \phi_n) = 1.0$. Moreover, over $(\mathcal{T},p')$,
we can rewrite the marginal probability of a query answer $\psi$ as follows.
\[
\begin{array}{rl}
P(\psi) \!\!\!\! &
\!\stackrel{\eqref{eq:lineageConf}}{=} \sum_{\mathcal{W} \subseteq \mathcal{T}, \mathcal \models \psi} P(\mathcal{W},\mathcal{T}) \\
& = \sum_{\mathcal{W} \subseteq \mathcal{T}, \mathcal{W} \models \psi, \mathcal{W} \models \phi_1, \dots , \mathcal{W} \models \phi_n} P(\mathcal{W},\mathcal{T}) \\
& = P(\psi \land \phi_1 \land \dots \land \phi_n) \\
\end{array}
\]
By combining both equations, we obtain over $(\mathcal{T},p')$:
\[
P(\psi~|~\phi_1 \land \dots \land \phi_n) = \frac{P(\psi \land \phi_1 \land \dots \land \phi_n)}{P(\phi_1 \land \dots \land \phi_n)} = \frac{P(\psi)}{1.0} = P(\psi)
\]
%\qedhere
\end{proof}
Thus, the learning problem subsumes conditioning PDBs~\cite{KochConditioningPDB}.
From a Bayesian perspective, the solution to the learning problem $p'$ can be seen as posterior probabilities of the base tuples in $\mathcal{T}_l$. 

\noindent\textbf{Learning by Conditioning.}
Following Definition~\ref{def:logicalObjective}'s logical objective,
we can apply constraint-enforcing
approaches %, like~\cite{DataIntegrationUserFeedback,KochConditioningPDB},
to solve a subset of possible learning problem instances. The subset is characterized
by instances with consistent labels, having $\mathcal{T} = \mathcal{T}_l$, and by restricting the lineage labels
to $l_i \in \{0.0, 1.0\}$.
We create a single constraint in the form of Equation~\eqref{eq:booleanOnly}'s conjunction, initially
set all tuple confidences to $0.5$, and solve the resulting conditioning problem~\cite{KochConditioningPDB}.

\subsection{Updating \& Cleaning PDBs} %Probabilistic Data\-bases}
\label{sec:updating}
\noindent \textbf{Updating.}
If we are given an existing probabilistic database $(\mathcal{T},p)$
and knowledge in the form of labeled lineage formulas
$\mathcal{L}:=\langle(\phi_1, l_1),\dots,(\phi_n,l_n)\rangle$,
we can update the tuples' probability values via the 
learning problem.
%(inserting, deleting, and updating tuples~\cite{Sarma:2010:LLV:1876037.1876072} is different).
We produce a new probabilistic database
$(\mathcal{T},p')$, whose probability values $p'$ are updated according
to the information provided in $\mathcal{L}$.
To achieve this, we create a learning problem instance (whose solution is $p'$)
by using $\mathcal{L}$, setting $\mathcal{T}_l := \mathcal{T}$ and 
defining a prior $\mathit{prior}(t) := p(t)$.

\noindent \textbf{Cleaning.}
The new probability values $p'$ allow for cleaning the probabilistic
database as follows. If $p'$ defines a tuple's
probability value to be $0.0$, we can delete it from the database. Conversely,
if $p'$ yields $1.0$ for a tuple's probability value, we can move it into
a new, deterministic relation. % that replaces the former probabilistic relation.

%For both the above updating and cleaning tasks, the resulting database is conditioned onto the lineage labels. Updating preserves all tuples (with their new probability values), while cleaning retains only those tuples with a learned probability value of 1.0. 

\subsection{Incomplete Databases}
\label{sec:incompleteAndLearning}
A field that is related to PDBs are incomplete databases. Intuitively, in an incomplete database, some attributes values or entire tuples may be missing in the given database instance. 
A completion of an incomplete database can be seen as a possible world in a PDB---with a probability.

%\subsubsection{Missing Attribute Values}
%\label{sec:missingAttributes}
\noindent\textbf{Missing Attribute Values.}
In~\cite{DerivingPDBs}, a PDB is derived
from an incomplete database which exhibits missing attributes
in some of its tuples.
Their idea is to estimate the probability of a possible completion of an
incomplete tuple from the complete part of the database.
We show that this approach is an instance
of the learning problem via the following reduction. 

%\noindent \textbf{Reduction.}
Let an incomplete database
be given by a set of complete tuples $\mathcal{T}_c$ and a set of incomplete
tuples $\mathcal{T}_i$.
We consider an incomplete tuple $R(\bar{a}) \in \mathcal{T}_i$ of relation $R$,
where one or more attributes in $\bar{a}$ lack values,
such that all possible completions are represented by $\bar{a}_i \supsetneq \bar{a}$
(assuming finite domains).
Then, we create a new uncertain relation $R' := \{\bar{a}_i~|~\bar{a}_i \supsetneq \bar{a}\}$
and add one deduction rule per completion $\bar{a}_i$:
\[
R(\bar{a}_i) \leftarrow R'(\bar{a}_i) \land \bigwedge_{j\neq i} \neg R'(\bar{a}_j)
\]
The above rules allow at most one completion of $\bar{a}$ to be true within a possible world, so the resulting lineage formulas form a block-independent PDB~\cite{PDBbook}.
Now, we create labels following~\cite{DerivingPDBs}'s approach.
For a subset of argument values $\bar{s} \subset \bar{a}$, we count how often
the complete tuples $\mathcal{T}_c$ feature the completion $\bar{a}_i$, in symbols
$
I_{\bar{s}}(\bar{a}_i) := |\{R(\bar{a}') \in \mathcal{T}_c~|~\bar{a} \cap \bar{a}' =\bar{s}, \bar{a}_i \backslash \bar{a} \subset \bar{a}'\}|
$.
Then, for each completion $\bar{a}_i$, we generate the label
$(R(\bar{a}_i),$ $\frac{I_{\bar{s}}(\bar{a}_i)}{\sum_j I_{\bar{s}}(\bar{a}_j)})$.
Besides these labels, the resulting learning problem instance uses
the new relations $R'$'s tuples in $\mathcal{T}_l$ .

%\subsubsection{Missing Tuples}
%\label{sec:missingTuples}
\noindent\textbf{Missing Tuples.} Generally, any database instance can be seen
as a finite subset of the crossproduct of its attributes' domains. We now
consider an incomplete database, whose (finite sets of) existing tuples
and potentially missing tuples are $\mathcal{T}_c$ and $\mathcal{T}_m$, respectively.
Assume we intend to enforce logical
%Generally, any database instance can be seen as a finite subset of the crossproduct of its attributes' domains.
%From a PDB perspective, it holds that $\forall t_c \in \mathcal{T}_c: p(t_c) = 1.0$
%and $\forall t_m \in \mathcal{T}_m: p(t_m) = 0.0$. Now, we intend to enforce logical
formulas $\phi_1, \dots, \phi_n$ over tuples $\mathcal{T}_c \cup \mathcal{T}_m$,
which could for example result from constraints or user feedback.
We create a learning problem instance by
setting $\mathcal{T} := \mathcal{T}_c \cup \mathcal{T}_m$,
$\mathcal{T}_l := \mathcal{T}_m$
and $\mathcal{L} := \langle(\phi_1, 1.0),\dots,(\phi_n, 1.0)\rangle$.
Thus, a solution to the learning problem will complete $\mathcal{T}_c$ with
(possibly uncertain) tuples from $\mathcal{T}_m$, such that the logical
formulas $\phi_1, \dots, \phi_n$ are fulfilled.

%% file: experiments.tex
\section{Experiments}
\label{sec:experiments}
Our evaluation focuses on the following four aspects.
First, we compare the quality of our approach
to learning techniques in SRL
(Section~\ref{sec:exp:uw-cse}) and to constraint-based reasoning techniques applied in information
extraction settings (Section~\ref{sec:exp:tfacts}). Second, we
compare the runtime behavior of our algorithm to SRL methods
(Section~\ref{sec:exp:scalingVSsrl}) and to other gradient-based optimization techniques
(Section~\ref{sec:exp:gradientBased}).
Third, we explore the scalability of
our method to large data sets (Section~\ref{sec:exp:scalability}).
Finally, in Section~\ref{sec:exp:objectives}, we investigate the runtime behavior
of the two objective functions defined in Section~\ref{sec:solving}.
Due to space constraints, additional experiments on
varying $\epsilon_\mathit{abs}$, $\epsilon_\mathit{rel}$
and Algorithm~\ref{alg:sgd}'s ability
to find global optima are available in the supplementary
material\footnote{\scriptsize{\label{footnote:supplementary}\href{http://www.mpi-inf.mpg.de/~mdylla/learning.pdf}{http://www.mpi-inf.mpg.de/\textasciitilde mdylla/learning.pdf}}}.

\begin{figure*}[ht]
\begin{center}
 \subfigure[Dataset Statistics]{ \label{table:datasets}
    \begin{small}
		$
		\begin{array}{c|l|c|r|r|r|c|c|c}
		\hline
		\text{\textbf{Section}} & \text{\textbf{Figure}} & \text{\textbf{Source}}
		  & |\mathcal{T}| & |\mathcal{T}_l| & |\mathcal{L}| & \text{Avg.}~T(\phi)
		  & \text{Boolean} & \text{Inconsistent} \\
		\hline
		\hline
		  \ref{sec:exp:uw-cse} & \ref{fig:srlTask} & \text{UW-CSE}\textsuperscript{\ref{footnote:uwcse}} & 2,161 & 49 & 113~\text{to}~452 & 5.8~\text{to}~8.3 & \text{yes} & \text{yes}\\
		\hline
		  \ref{sec:exp:tfacts} & \ref{fig:pravda} & \text{PRAVDA}\textsuperscript{\ref{footnote:pravda}}& 75,091 & 37,383 & 89,874 & 2.3 & \text{no} & \text{yes}\\
		\hline
      \ref{sec:exp:scalingVSsrl} & \ref{fig:smallScaleTask} & \text{synthetic} & 100 & 100 & 10~\text{to}~100 & 5.8 & \text{yes} & \text{some} \\
		\hline
			\ref{sec:exp:gradientBased} & \ref{fig:gradientBased} & \text{YAGO2}\textsuperscript{\ref{footnote:yago}} & 224, 440, 854 & 19,985 & 5,562 & 3.6 & \text{no} & \text{no} \\
		\hline
      \ref{sec:exp:scalability} & \ref{fig:largeScaleTask}~P1 & & & 217,846 & 228,050 & 2.7  & \text{no} & \text{yes} \\
      \ref{sec:exp:scalability} & \ref{fig:largeScaleTask}~P2 & \text{YAGO2}\textsuperscript{\ref{footnote:yago}} & 224, 440, 854 & 217,846 & 79,600 & 60.6  & \text{no} & \text{yes} \\
      \ref{sec:exp:scalability} & \ref{fig:largeScaleTask}~P3 & & & 1,721,156 & 459,597 & 3.7 & \text{no} & \text{no} \\
		\hline
      \ref{sec:exp:objectives} & \ref{fig:objectives} & \text{synthetic} & 100 & 100 & 1~\text{to}~15 & 5.8 & \text{yes} & \text{no} \\
      \hline
		\end{array}
		$
		\end{small}
	}
 \subfigure[Quality Task: SRL Data]{ \label{fig:srlTask}
    \includegraphics[width=0.3\textwidth]{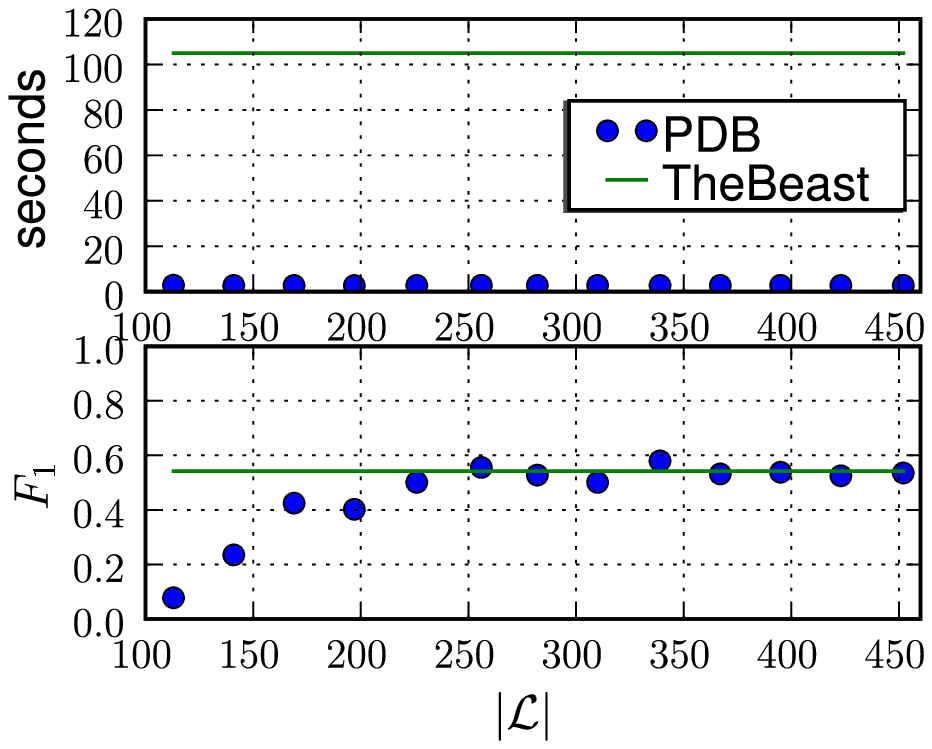}
 }
 \subfigure[Quality Task: Fact Extraction]{ \label{fig:pravda}
    \includegraphics[width=0.3\textwidth]{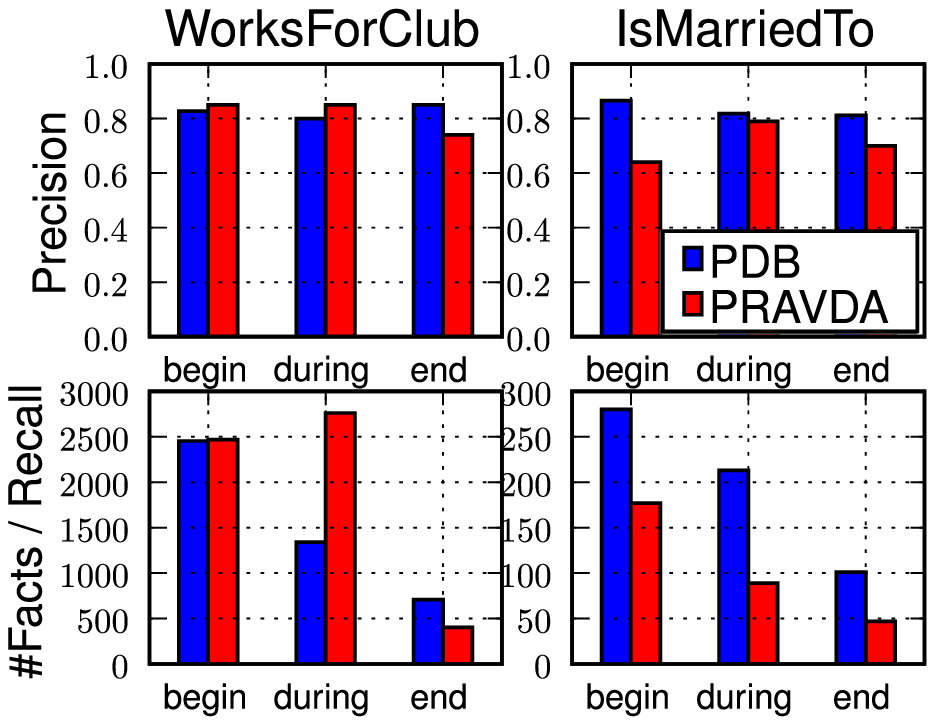}
 }
 \subfigure[Runtime Task: SRL Methods]{ \label{fig:smallScaleTask}
    \includegraphics[width=0.3\textwidth]{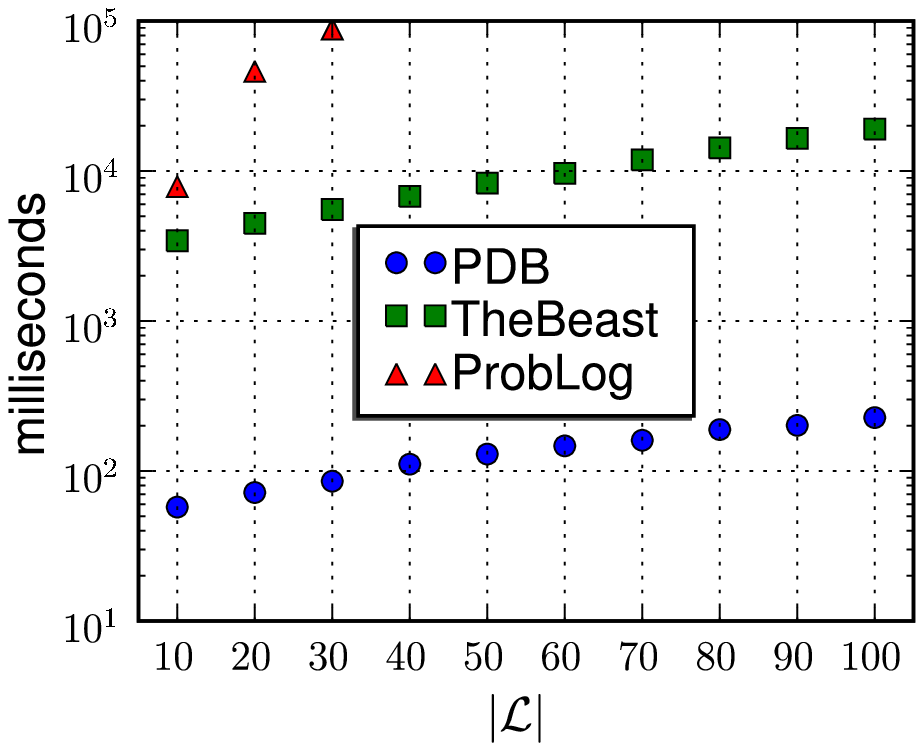}
 }
 \subfigure[Runtime Task: Gradient Methods]{ \label{fig:gradientBased}
 		\includegraphics[width=0.3\textwidth]{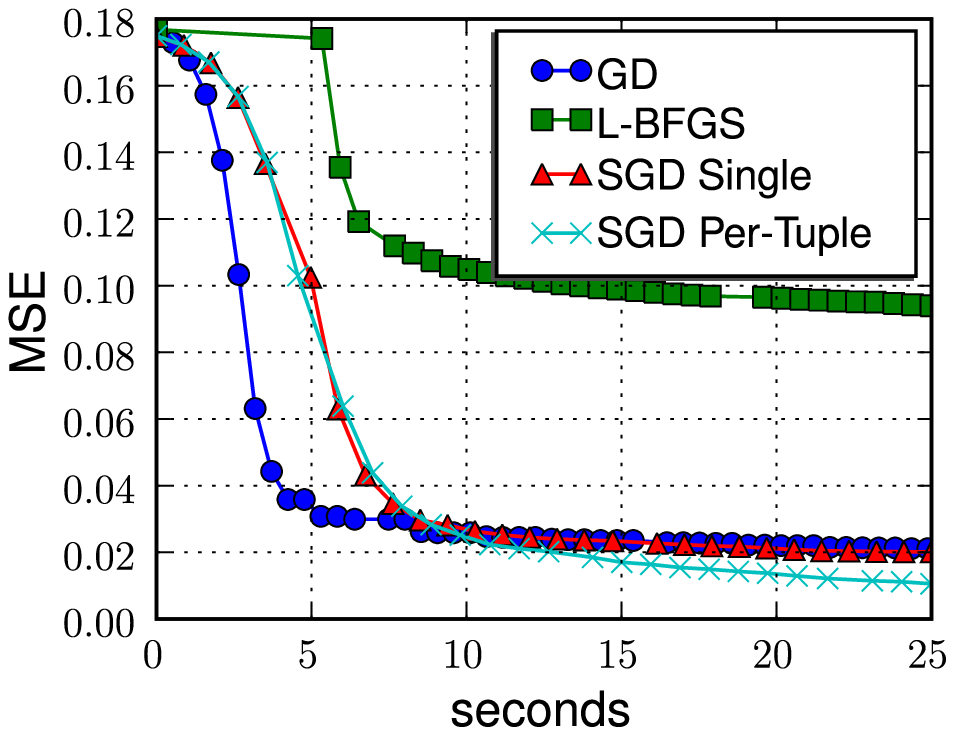}
 }
 \subfigure[Runtime Task: Large Scale]{ \label{fig:largeScaleTask}
    \includegraphics[width=0.3\textwidth]{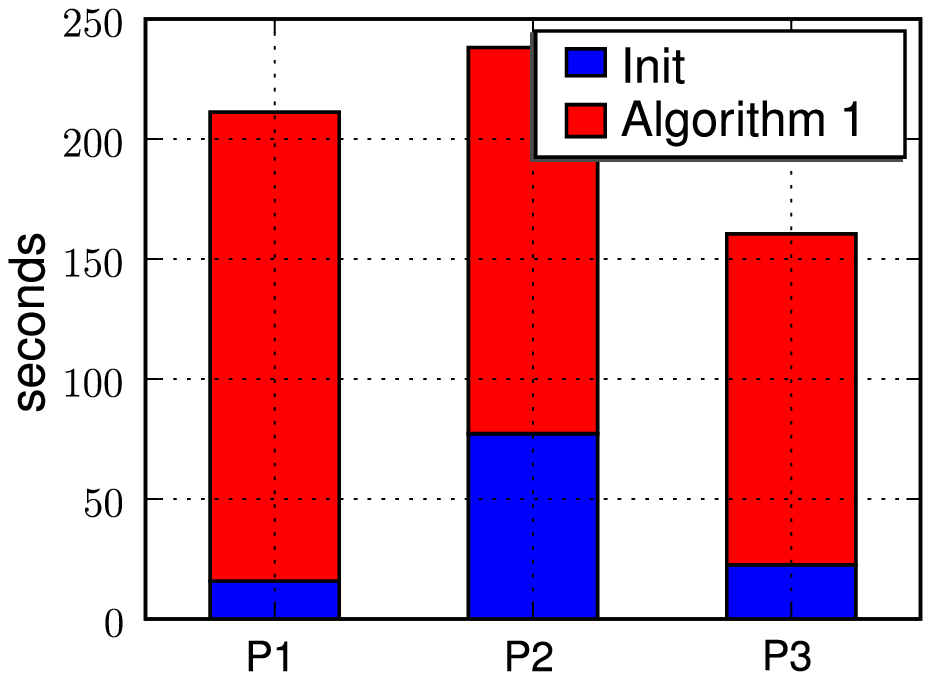}
 }
 \subfigure[Runtime Task: Objectives]{ \label{fig:objectives}
    \includegraphics[width=0.3\textwidth]{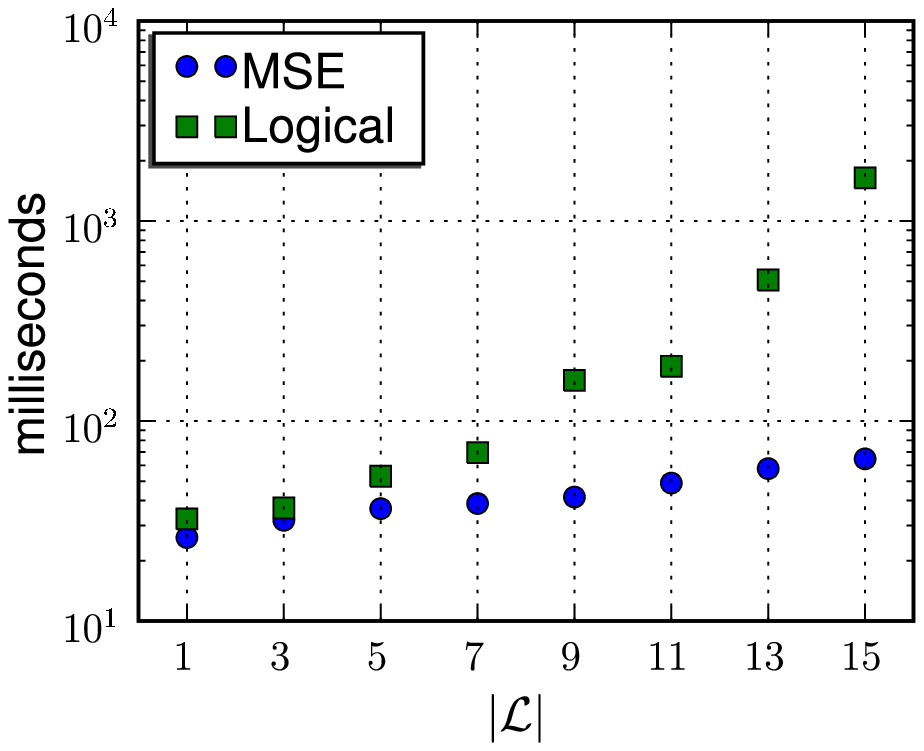}
 }
\vspace{-4mm}
\caption{Experiments} \label{fig:experiments}\end{center}
\vspace{-6mm}
\end{figure*}

\noindent \textbf{Overview.} As an overview, we present the basic characteristics of all learning
problem instances in Figure~\ref{table:datasets}, where {\em Avg.\ $T(\phi)$} is calculated
as $\frac{1}{|\mathcal{L}|} \sum_{(\phi_i, l_i) \in \mathcal{L}} |T(\phi_i)|$.

\noindent \textbf{Setup.}
Our engine %, whose source code we publish with this paper\textsuperscript{\ref{footnote:sourceCode}}, 
is implemented in Java. It employs a PostgreSQL 8.4 database backend for evaluating Datalog rules in a bottom-up manner and to instantiate lineage
formulas. 
If not stated otherwise, {\em PDB} refers to Algorithm~\ref{alg:sgd}'s
implementation with the MSE objective and a per-tuple learning rate. 
For checking convergence, we set $\epsilon_\mathit{abs} = 10^{-6}$
and $\epsilon_\mathit{rel} = 10^{-4}$. 
We ran all experiments on an 8-core Intel Xeon 2.4GHz machine
with 48 GB RAM, repeated each setting four times, and report the
average of the last three runs. 
Whenever different programs compete,
all of them run in single-threaded mode.
All rules used in the experiments
are provided in the supplementary material\textsuperscript{\ref{footnote:supplementary}}.

\vspace{-1mm}
\subsection{Quality Task: SRL Setting}
\label{sec:exp:uw-cse}
%In this task, we compare the predictive performance a PDB with tuple
%confidence learning to established methods from SRL. 
\noindent\textbf{Dataset.}
We use the openly available UW-CSE
dataset\footnote{\scriptsize{\label{footnote:uwcse}\href{http://alchemy.cs.washington.edu/data/uw-cse/}{http://alchemy.cs.washington.edu/data/uw-cse/}}},
which comprises a database describing the
University of Washington's computer science department via
the following relations:
{\em AdvisedBy}, {\em CourseLevel}, {\em HasPosition}, {\em InPhase},
{\em Professor}, {\em ProjectMember}, {\em Publication}, {\em Student},
{\em TaughtBy}, {\em Ta} (teaching assistant), and {\em YearsInProgram}.
Moreover, the dataset is split into five sub-departments, and
%namely artificial intelligence, graphics, language, systems, and theory.
we consider this dataset's relations to be deterministic. %certain, i.e.\ non-probabilistic.

\noindent\textbf{Task.}
The goal is inspired by an experiment in~\cite{MLN}, namely 
to predict the {\em AdvisedBy} relation from all input
relations except {\em Student} and {\em Professor}. We train and test
in a leave-one-out fashion by sub-department.

\noindent\textbf{Rules.}
We automatically
create 49 rules resembling all joins (including self-joins)
between two relations (except {\em student}, {\em professor}, and
{\em AdvisedBy}), having at least one argument of type person.
Furthermore, we add one uncertain relation {\em rules}, containing one
tuple for each of the 49 rules and include the corresponding tuple in the join,
for example:
\vspace{-1mm}
\[
\mathit{AdvisedBy}(P_1,P_2) \leftarrow
\left(
\!\!\!
\begin{array}{r}
\mathit{Ta}(D,C,P_1,T) \land \\
\mathit{TaughtBy}(D,C,P_2,T) \land \\
\end{array}
\!\!\!\mathit{Rules}(1) \right)
\vspace{-1mm}
\]
The remaining rules are given in the
supplementary material\textsuperscript{\ref{footnote:supplementary}}.
%By this construction, all lineage instances of a single
%rule share the same uncertain tuple (e.g., $\mathit{rules}(1)$).
We learn the probability values of the 49 tuples, hence classifying how well each 
rule predicts the $\mathit{AdvisedBy}$ relation.
%A more elaborated discussion on this model of uncertain rules is
%available in~\cite{DyllaMiliarakiTheobald2012}.

\noindent \textbf{Labels.}
Regarding labels, we used the 113 instances of {\em AdvisedBy} as {\em positive labels},
i.e., all their probability labels are $1.0$.
In addition, there are about 16,000 
person-person pairs not contained in {\em AdvisedBy}.
%Adhering the closed world assumption 
We randomly draw pairs from these as {\em negative labels} (with a probability label of $0.0$).

\noindent\textbf{SRL Competitors.}
We compete with {\em TheBeast}~\cite{riedel08}, the fastest Markov
Logic~\cite{MLN} implementation we are aware of.
It uses an 
in-memory database and performs inference via Integer Linear Programming.
We ran it on the
same set of data and rules. % and confidences to alter. 
Additionally, we ran the probabilistic Prolog engine 
{\em ProbLog}~\cite{ProbLogLearning}, but even on the reduced
datasize of one sub-department it did not terminate after one hour.

\noindent\textbf{Results.}
In Figure~\ref{fig:srlTask}, we depict both the runtimes as well as the
prediction quality in terms of the $F_1$ measure (the harmonic mean of precision
and recall) for the $\mathit{AdvisedBy}$ relation. {\em TheBeast} is
a straight line, since it allows only positive labels.
%, such that it does not get affected by adding negative examples.
For {\em PDB}, we started with all positive labels and added increasing
numbers of negative labels.

\noindent\textbf{Analysis.}
Regarding runtimes, {\em PDB} is consistently about 40 times faster
than {\em TheBeast}. With respect to $F_1$, adding more negative labels
to {\em PDB} yields improvements until we saturate at the same level
as {\em TheBeast}.

\vspace{-1mm}
\subsection{Quality Task: Information Extraction}
\label{sec:exp:tfacts}
%In the spirit of Example~\ref{ex:intro}, we tackle the problem
%of extracting facts from
%free text by the use of textual patterns. In addition, we consider facts with 
%temporal annotations in the form of time intervals specifying the facts'
%validity as described in~\cite{ACL12yafang} .

\noindent\textbf{Dataset.}
This dataset\footnote{\scriptsize{\label{footnote:pravda}
\href{http://www.mpi-inf.mpg.de/yago-naga/pravda/}{http://www.mpi-inf.mpg.de/yago-naga/pravda/}}}
contains about 450,000 crawled web pages in
the sports and celebrities domains, where
about 12,500 textual patterns are used to extract facts.

\noindent\textbf{Task.}
Following~\cite{ACL12yafang}, we consider two different relations,
namely {\em WorksForClub} in the
sports domain and {\em IsMarriedTo} in the celebrities domain.
Both relations contain facts with 
temporal annotations. %, which specify time intervals during which the facts are valid. % as described in~\cite{ACL12yafang} .
The goal is to determine, for each textual pattern, whether it expresses
a temporal {\em begin}, {\em during} or {\em end} event of one of the two
relations, or none of them.
For example, for {\em WorksForClub}, we could find that David Beckham joined Real Madrid in 2003
{\em (begin)}, scored goals for them in 2005 {\em (during)}, and left the club in 2007
{\em (end)}.
%Having classified the patterns, occurrences of entity pairs next to a 
%textual pattern in the input documents thus yield the facts.

\noindent\textbf{PDB Setup.}
We model temporal data in the PDB according to~\cite{TPDBdylla}.
Text occurrences of a potential fact are stored in the deterministic relation
$\mathit{Occurrence}(\mathit{Pid}, \mathit{E1}, \mathit{E2}, \mathit{Types},$ $\mathit{Begin}, \mathit{End})$,
where {\em Types} holds the
entities' types and {\em Begin}, {\em End}
contain integers encoding the limits of their occurrences' time intervals.
To encode the decision whether a pattern expresses a temporal {\em begin},
{\em during}, or {\em end} event, we instantiate three uncertain relations
$\mathit{Begin}(\mathit{Pid})$, $\mathit{During}(\mathit{Pid})$,
and $\mathit{End}(\mathit{Pid})$, which each hold one entry per pattern
and whose probability values we learn.
Text occurrences of potential facts are connected to the patterns
by six rules (see\textsuperscript{\ref{footnote:supplementary}} for details)
of the following kind
\vspace{-1mm}
\[
\begin{array}{l}
\mathit{IsMarriedToBegin}(E_1,E_2,T_1,T_2) \\
~~\leftarrow \mathit{Begin}(\mathit{Pid}) \land \mathit{Occurrence}(\mathit{Pid},E_1,E_1,pp,T_1,T_2) \\
\end{array}
\vspace{-1mm}
\]
where $pp$ stands for person-person type pair. 
To enforce that a textual pattern expresses at most one of {\em begin},
{\em during}, or {\em end}, we make them mutually exclusive via the rules
\vspace{-1mm}
\[
\begin{array}{rcl}
\mathit{Constraint1}(\mathit{Pid}) & \leftarrow & \mathit{Begin}(\mathit{Pid})\land \mathit{During}(\mathit{Pid}) \\
\mathit{Constraint2}(\mathit{Pid}) & \leftarrow & \mathit{Begin}(\mathit{Pid})\land \mathit{End}(\mathit{Pid}) \\
\mathit{Constraint3}(\mathit{Pid}) & \leftarrow & \mathit{During}(\mathit{Pid})\land \mathit{End}(\mathit{Pid}) \\
\end{array}
\vspace{-1mm}
\]
whose resulting lineage formulas we label with $0.0$.
Moreover, we use temporal precedence constraints by instantiating
six rules of the form
\vspace{-1mm}
\[
\mathit{Constraint4}(E_1,E_2)
\leftarrow \!\!\!
\begin{array}{l}
\mathit{IsMarriedToBegin}(E_1,E_2,T_1,T_2) \land \\
\mathit{IsMarriedToDuring}(E_1,E_2,T_3,T_4) \\
~~~\land T_3 < T_2 \\
\end{array}
\vspace{-1mm}
\]
and label their lineage with $0.0$.
Finally, we employ the 266 labels for textual patterns
and the 341 labels for facts from the original work~\cite{ACL12yafang}.

\noindent\textbf{Competitor.}
The authors of~\cite{ACL12yafang} utilized a combination of Label Propagation and Integer Linear Programming to rate the
textual patterns and to enforce temporal constraints. %, respectively.

\noindent\textbf{Results.}
In Figure~\ref{fig:pravda}, we report our system's {\em (PDB)} result
along with the best result from~\cite{ACL12yafang} {\em (PRAVDA)}.
To evaluate precision, we sampled 100 facts per relation and event type
and annotated them manually. Recall is the absolute number of facts obtained.

\noindent\textbf{Analysis.}
For relations with a few, decisive textual patterns, {\em PDB} keeps
up with precision, while slightly gaining in recall, probably due
to the relaxation of constraints by the MSE objective. However, for {\em worksForClub}'s
{\em during} relation, there is a vast number of relevant patterns, which
puts Label Propagation's undirected model in favor, whereas our directed model
suffers in terms of recall.

\vspace{-1mm}
\subsection{Runtime Task: SRL Methods}
\label{sec:exp:scalingVSsrl}
%We compare the runtimes of SRL implementations and
%our PDB implementation with learning.
\noindent \textbf{Setup.}
To systematically verify scalability, we create synthetic
data sets as follows. We fix $\mathcal{T} = \mathcal{T}_l$ to 100 tuples.
Then, we instantiate a growing number of lineage formulas of the form
$(t \land \neg t \land \neg t) \lor (t \land \neg t \land \neg t)$,
where all tuple identifiers are uniformly drawn from $\mathcal{T}_l$,
and negations exist with probability $0.5$. Each formula's probability label
is randomly set to either $0.0$ or $1.0$.

\noindent \textbf{Competitors.} 
Besides {\em TheBeast}~\cite{riedel08}, we compete with
{\em ProbLog}~\cite{ProbLogLearning}, a probabilistic Prolog engine, whose grounding techniques and distribution semantics are closest
to ours.

\noindent \textbf{Results.}
For each value of $|\mathcal{L}|$, we create five problem instances and
depict their average runtime in Figure~\ref{fig:smallScaleTask}.

\noindent \textbf{Analysis.} 
{\em PDB} converges on average about 600 times faster
than {\em ProbLog} and about 70 times faster than {\em TheBeast}.

\vspace{-1mm}
\subsection{Runtime Task: Gradient Methods}
\label{sec:exp:gradientBased}
%We compare Algorithm~\ref{alg:sgd}'s stochastic gradient
%descent (SGD) with per-tuple learning rate to other
%gradient-based optimization methods.
\noindent \textbf{Setup.}
We employ the openly available
YAGO2\footnote{\scriptsize{\label{footnote:yago}\href{http://www.mpi-inf.mpg.de/yago-naga/yago/}{http://www.mpi-inf.mpg.de/yago-naga/yago/}}} knowledge base, which comprises
about 110 relations.
The task is to learn the probability values of tuples $\mathcal{T}_l$ in the {\em LivesIn} relation.
Moreover, we label the following rule's
\vspace{-1mm}
\[
\mathit{ToLabel}(L) \leftarrow \mathit{LivesIn}(P,L)
\vspace{-1mm}
\]
lineage formulas with synthetic target probabilities.
Since the rule's projection on the first argument makes all lineage
formulas $\phi$ disjoint with respect to their tuples $T(\phi)$, the resulting learning problem
instance is consistent. Hence, its global optima have a mean
squared error (MSE) of $0.0$.

\noindent \textbf{Competitors.}
Algorithm~\ref{alg:sgd} with per tuple learning rate
{\em (SGD Per-Tuple)} competes with
a single learning rate {\em (SGD Single)},
with gradient descent {\em (GD)}, and with {\em L-BFGS}~\cite{OptimizationBook},
which approximates the Hessian with its second derivatives. All methods
are initialized with the same learning rate.

\noindent \textbf{Results.}
We plot the MSE against the runtime of the different methods in Figure~\ref{fig:gradientBased}.

\noindent \textbf{Analysis.}
{\em GD} takes less time per iteration.
Hence its curves drops faster in the beginning, but then stagnates.
The two {\em SGD} variants behave similarly at first. Later on,
the per-tuple learning rate yields constant improvements, whereas
the single learning rate does not.
{\em L-BFGS}, finally, improves slowly in comparison.

\vspace{-1mm}
\subsection{Runtime Task: Scalability}
\label{sec:exp:scalability}
%In this experiment we study the scalability of our implementation.
\noindent \textbf{Dataset.} As previously,
we run on YAGO2\textsuperscript{\ref{footnote:yago}}.
For tuples in $\mathcal{T} \backslash \mathcal{T}_l$,
we use uniformly drawn synthetic probability values.

\noindent \textbf{Labels.}
In order to create labels, we run queries on YAGO2
and label their answers' lineage formulas with
synthetic target probabilities (see\textsuperscript{\ref{footnote:supplementary}} for details).

\noindent \textbf{Results.}
Figure~\ref{fig:largeScaleTask} contains the results of
three large learning problem instances $P1$ to $P3$, where
{\em Init} is the time spent on instantiating the
lineage formulas, and {\em Algorithm~\ref{alg:sgd}} had multi-threading enabled.

\noindent \textbf{Analysis.} 
The {\em Init} time is determined by the number ($|\mathcal{L}|$)
and size {\em (Avg. $T(\phi)$)} of lineage formulas being instantiated.
{\em Algorithm~\ref{alg:sgd}} is faster on consistent instances ($P3$).
Its runtime is dominated by the number of labels per tuple $t_l \in \mathcal{T}_l$.

\vspace{-1mm}
\subsection{Runtime Task: Objectives}
\label{sec:exp:objectives}
\noindent \textbf{Setup.}
As a last experiment, we run Algorithm~\ref{alg:sgd} once
with the {\em Logical} objective (see Definition~\ref{def:logicalObjective})
and once with the mean-squared-error {\em (MSE)} objective
(see Definition~\ref{def:MSE}).
The synthetic data is created analogously to Section~\ref{sec:exp:scalingVSsrl}.

\noindent \textbf{Results \& Analysis.}
Already on tiny instances of up to 15 labels as in Figure~\ref{fig:objectives},
the {\em Logical} objective
slows down significantly in comparison to {\em MSE}, due to the expensive marginal computations of Equation~\eqref{eq:booleanOnly}.

%% file: conclusions.tex
\section{Conclusions}
We introduced a novel method for learning tuple confidences in tuple-independent
probabilistic databases. We analyzed the properties of this learning problem from a theoretical perspective, devised gradient-based solutions, investigated the relationship to other problems, and
presented an implementation together with extensive experiments.
For future work, we see numerous promising directions.
Studying tractable subclasses of the learning problem
or dropping the tuple-independence assumption would improve our
theoretical understanding. Other valuable
targets lie in the creation of a large, publicly available benchmark and the
application of the learning problem to a broader range of related problems,
e.g., inspired by the ones mentioned in Section~\ref{sec:relatedProblems}.

%in final camera-ready version only
%\noindent\textbf{Acknowledgments.} The authors would like to thank Radu Curticapean for
%pointing out Bezout's theorem. Also, Rainer Gemulla's knowledge on stochastic gradient
%descent was very helpfull.